\documentclass[sn-basic]{sn-jnl}% Basic Springer Nature Reference Style/Chemistry Reference Style

\jyear{2023}%

\theoremstyle{thmstyleone}%

\theoremstyle{thmstyletwo}%

\theoremstyle{thmstylethree}%

\begin{document}

\title[Cool Star Magnetism]{Scaling and Evolution of Stellar Magnetic Activity}

\author[1,2]{\fnm{Emre} \sur{I\c{s}{\i}k}}\email{isik@mps.mpg.de}
\equalcont{These authors contributed equally to this work.}

\author[3]{\fnm{Jennifer L.} \sur{van Saders}}\email{jlvs@hawaii.edu}
\equalcont{These authors contributed equally to this work.}

\author*[4]{\fnm{Ansgar} \sur{Reiners}}\email{Ansgar.Reiners@phys.uni-goettingen.de}
\equalcont{These authors contributed equally to this work.}

\author[5]{\fnm{Travis S.} \sur{Metcalfe}}\email{travis@wdrc.org}

\equalcont{These authors contributed equally to this work.}

\affil[1]{\orgname{Max-Planck-Institut f\"ur Sonnensystemforschung}, \orgaddress{\street{Justus-von-Liebig-Weg 3}, \city{G\"ottingen}, \postcode{37077}, \country{Germany}}}

\affil[2]{\orgdiv{Department of Computer Science}, \orgname{Turkish-German 
University}, \orgaddress{\street{\c{S}ahinkaya Cd. 94}, \city{Beykoz}, \postcode{34820}, \state{Istanbul}, \country{Turkey}}}

\affil[3]{\orgdiv{Institute for Astronomy}, \orgname{University of Hawaii}, \orgaddress{2680 Woodlawn Dr.}, \city{Honolulu}, \state{Hawaii} \postcode{96822}, \country{USA}}

\affil*[4]{\orgdiv{Institut f\"ur Astrophysik und Geophysik}, \orgname{Georg-August-Universit\"at G\"ottingen}, \orgaddress{\street{Friedrich-Hund-Platz 1}, \city{G\"ottingen}, \postcode{37077}, \country{Germany}}}

\affil[5]{\orgname{White Dwarf Research Corporation}, \orgaddress{\street{9020 Brumm Trl}, \city{Golden}, \state{Colorado} \postcode{80403}, \country{USA}}}

\abstract{
Magnetic activity is a ubiquitous feature of stars with convective outer 
layers, with implications from stellar evolution to planetary atmospheres. 
Investigating the mechanisms responsible for the
observed stellar activity signals from days to billions of years is important in 
deepening our understanding of the spatial configurations and temporal patterns 
of stellar dynamos, including that of the Sun. In this paper, we focus on 
three problems and their possible solutions. We start with direct 
field measurements and show how they probe the dependence of magnetic 
flux and its density on stellar properties and activity indicators. 
Next, we review the current state-of-the-art in physics-based models of photospheric 
activity patterns and their variation from rotational to activity-cycle timescales. 
We then outline the current state of understanding in the long-term evolution of 
stellar dynamos, first by using chromospheric and coronal activity diagnostics, 
then with model-based implications on magnetic braking, which is the key mechanism 
by which stars spin down and become inactive as they age. We conclude by discussing possible 
directions to improve the modeling and analysis of stellar magnetic fields. 
}

\keywords{Cool Stars, Stellar Magnetism, Stellar Activity, Angular Momentum Loss}

\maketitle

\section{Overview}\label{sec1}

Magnetism is ubiquitous in stars and yet it is relatively poorly understood, even in our closest neighbor. We necessarily rely on observational constraints---direct measurements or proxies for magnetism---to probe magnetic behavior across stellar types and lifetimes, and to connect these observations to underlying theoretical descriptions \citep{schrijver00}. In this paper, we highlight ways in which magnetism on stars reveals itself, and the insights those physical manifestations provide about the underlying physics of magnetic fields in stellar systems. 
The purpose of this paper is to report on recent progress in the following particular problems. 
\begin{itemize}
\item How does magnetic flux and its density scale with rotation and the fractional depth of the convection-zone? 
\item How can physics-based diagnostic modeling help us to constrain surface patterns and their evolution?
\item What is responsible for the spin-down and the weakening of outer-atmospheric activity indicators with age?
\end{itemize}
Following a summary of our recent attempts to find answers to these questions, we present an outlook on possible 
avenues to better understand the scaling relations of stellar magnetic activity. More extensive reviews 
can be found in the literature \citep[e.g.,][]{2009ARA&A..47..333D,Strassmeier09,2012LRSP....9....1R,Engvold19,Basri21}.

The structure and dynamics of magnetic fields threading the atmospheres 
of stars other than the Sun are observed mostly indirectly. Magnetic
field measurements from Zeeman splitting of photospheric 
spectral lines are becoming more accessible owing to instrumentation at optical and near-infrared wavelengths, such as CRIRES$^{+}$ \citep{2014Msngr.156....7D}, PEPSI \citep{2015AN....336..324S}, ESPaDOnS \citep{Donati+06}, NARVAL \citep{Auriere03}, HARPSpol \citep{Snik08,Piskunov11}, SPIRou \citep{Donati20}, and CARMENES \citep{carmenes} and HPF \citep{2012SPIE.8446E..1SM}. This promotes reliable quantification of the magnetic flux and its heating mechanisms observable in indirect activity indicators, and their scaling laws for different types of stars.
We cover recent work on direct magnetic field measurements and their use in constraining 
the rotation-activity scalings in Section~\ref{sec2}.

While our understanding is often driven by observations, numerical 
simulation frameworks are essential tools to
better evaluate observational trends of stellar magnetic activity 
on cool stars. Forward modeling of observational diagnostics are 
mostly based on physical models developed originally for 
the Sun. Scaling laws are often used to extend the solar paradigm 
to younger and more active suns as well as for cooler stars with 
deeper convection zones. More physically motivated applications 
involve dynamo models of the global magnetic 
field and the flux emergence process as a function of stellar 
properties. We present some important recent developments in 
modeling photospheric diagnostics of stellar magnetism in 
Section~\ref{sec3}. 

The indirect diagnostics---also called proxies---of magnetic 
activity include: (1) disk-integrated 
brightness in intermediate and broad bandpasses for the effects on the photosphere;
(2) narrow-band radiative fluxes 
centered on spectral lines such as 
H$\alpha$ and singly ionized Ca and Mg that 
probe the chromosphere; and 
(3) outer-atmospheric indicators of non-thermal heating by magnetic 
fields, at X-ray and radio wavelengths. 
We review recent advances in chromospheric and coronal proxies in 
relation to the long-term evolution of stellar activity in 
Section~\ref{sec4}. 

Stars with convective outer layers have dynamos 
that support large-scale fields from their interiors 
to their magnetospheric environments, also called astrospheres. 
Following star formation and disk dispersal, rapid stellar 
rotation coupled with convection leads to strong magnetic 
activity, which is responsible for strong magnetized winds removing 
angular momentum from the star. Rotational evolution of cool stars 
can be used as a proxy for the integrated large-scale field 
behavior. We discuss the connection between magnetic braking 
and global field properties in Section~\ref{sec5}. 

We propose 
strategies for further progress in the problems introduced above in Section~\ref{sec6}.

%\section{Direct Magnetic Field Measurements}\label{sec2}
\section{Scaling of magnetic flux and non-thermal emission}\label{sec2}

Our picture of stellar magnetism and its influence on stellar evolution and
activity is anchored in the detailed observational data from the
Sun. Among other examples, the spatial correspondence between active regions and magnetic flux concentrations, the occurrence of faculae and dark spots along latitudinal bands, differential rotation, the thermal and magnetic structure of active regions, and the temporal variability including magnetic cycle(s), are phenomena that can be observed in the Sun \citep[see, e.g.,][]{1987A&A...180..241S, 2003A&ARv..11..153S, 2006RPPh...69..563S}. In analogy, they are assumed to occur on other stars \citep[e.g.,][]{2005LRSP....2....8B}.

As a \emph{direct} measurement of magnetic fields, we understand determination of the immediate influence of the field, e.g., the signatures of the Zeeman effect on line profiles. This is in contrast to \emph{indirect} measurements, which include the measurement of proxies of magnetic activity, for example non-thermal emission. The direct measurement of magnetic fields is hampered by the fact that other stars are very far away and cannot be spatially resolved. One of the
consequences is that spectroscopy can be obtained from the spatially
integrated stars but not from individual areas, which leads to blending of spectroscopic effects caused by
magnetism with those from other atmospheric effects and partial
cancellation of polarization  \citep[see,
e.g.,][]{2009ARA&A..47..333D, 2012LRSP....9....1R,
2021A&ARv..29....1K}. Direct observations of magnetic fields therefore
typically require substantial spectral resolution and signal-to-noise (S/N),
and sometimes recurrent observations of the same star over different
rotational phases. Therefore, other \emph{indirect} indicators of stellar
magnetism are often employed to characterize stellar magnetic activity. These
are usually indicators of non-thermal emission that is generated by the
stellar magnetic field in analogy to the solar example \citep[see, e.g.,][and references therein]{2022A&A...662A..41R}.

Direct measurements of magnetism are usually those that investigate an
immediate observable of the magnetic fields, i.e., spectroscopic signatures
like the Zeeman effect. In solar-type and low-mass stars, the Zeeman effect
causes the most often used \emph{direct} signatures that are Zeeman broadening
\citep[observed in integrated light, Stokes~\emph{I};][]{1988ApJ...324..441S}
and polarization \citep[see, e.g.,][]{2004ASSL..307.....L}. Observational
biases can be rather strong leading to large uncertainties in the measured
field strengths and/or the amount of magnetism unseen by the
observations \citep[see, e.g.,][and references therein]{2020ApJ...902...43K} of
several directions of polarization, and monitoring projects delivering very
high S/N data have provided a wealth of information from direct field
measurements.

Solar-type and low-mass stars show a clear relation between rotation and
non-thermal emission observed in several activity indicators, e.g., X-rays
\citep{Skumanich1972, Noyes1984, Pizzolato2003,
Wright2011, 2014ApJ...794..144R}. The causal connection between
stellar rotation, stellar activity, and rotational evolution is supposed to be
(1) a mechanism providing more surface magnetic flux at high rotation rates in any given
star, and (2) the generation of non-thermal emission in proportion to magnetic
energy (or magnetic flux) at the stellar surface.
Currently, there is no physical model that explains any of the two phenomena from first principles. 
%%%%%%%%%%%%%%%%%%%%%%%%%%%%%%%%%%%%%%%%%%%%%%%%%%%%%%%%%%%%%%%%%%%%%%%%%%%%%%%%%%%%%%%
\begin{figure}
\centering\includegraphics[width=0.9\textwidth]{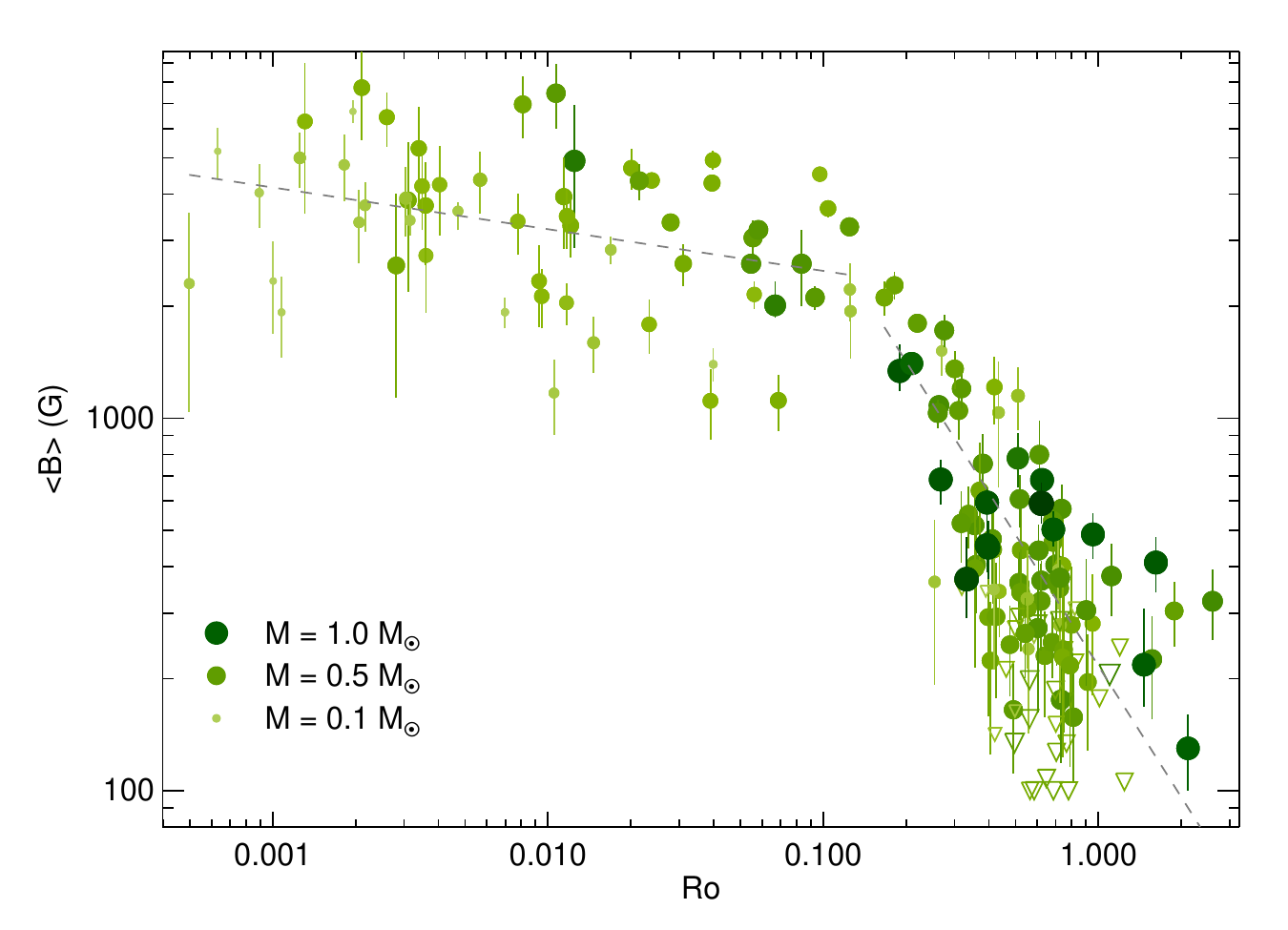}
\caption{Average surface magnetic field measurements from Stokes $I$ in sun-like and low-mass stars \citep[adapted from][]{2022A&A...662A..41R}. Grey dashed lines indicate the relation between field strength and Rossby number, $Ro = P/\tau$, for two groups of ``slow" and ``fast" rotators. Stellar mass is indicated with symbol size and color, downward triangles indicate upper limits in $<$$B$$>$. \label{fig:Bave_Ro}}
\end{figure}
%%%%%%%%%%%%%%%%%%%%%%%%%%%%%%%%%%%%%%%%%%%%%%%%%%%%%%%%%%%%%%%%%%%%%%%%%%%%%%%%%%%%%%%

The relation between average surface magnetic flux density (or field strength,
$\langle B\rangle$) and rotation is shown in Fig.\,\ref{fig:Bave_Ro}. Similar to the
rotation-activity relation, the Rossby number, $Ro = P/\tau$ with $\tau$ the
convective turnover time, is used as a normalized proxy of rotation. Fig.\,\ref{fig:Bave_Ro} shows two regimes of magnetic fields: a
group of rapid rotators with $Ro \le 0.1$ and $\langle B\rangle > 1$\,kG, and the slower
rotators with significantly weaker fields. Both groups show a statistically
significant relation between $\langle B\rangle$ and $Ro$. The dependence of $\langle B\rangle$ on $Ro$
is a lot stronger among the slow rotators than within the more rapidly
rotating group. The dashed lines show the relations:

\begin{eqnarray}
    \langle B \rangle\, & = & \phantom{2}199\,{\rm G} \times Ro^{-1.26 \pm 0.10} {\rm ~for~slow~rotation}\\
    \langle B \rangle\, & = & 2050\,{\rm G} \times Ro^{-0.11 \pm 0.03} {\rm ~for~rapid~rotation}
\end{eqnarray}

%%%%%%%%%%%%%%%%%%%%%%%%%%%%%%%%%%%%%%%%%%%%%%%%%%%%%%%%%%%%%%%%%%%%%%%%%%%%%%%%%%%%%%%
\begin{figure}
\centering\includegraphics[width=0.7\textwidth]{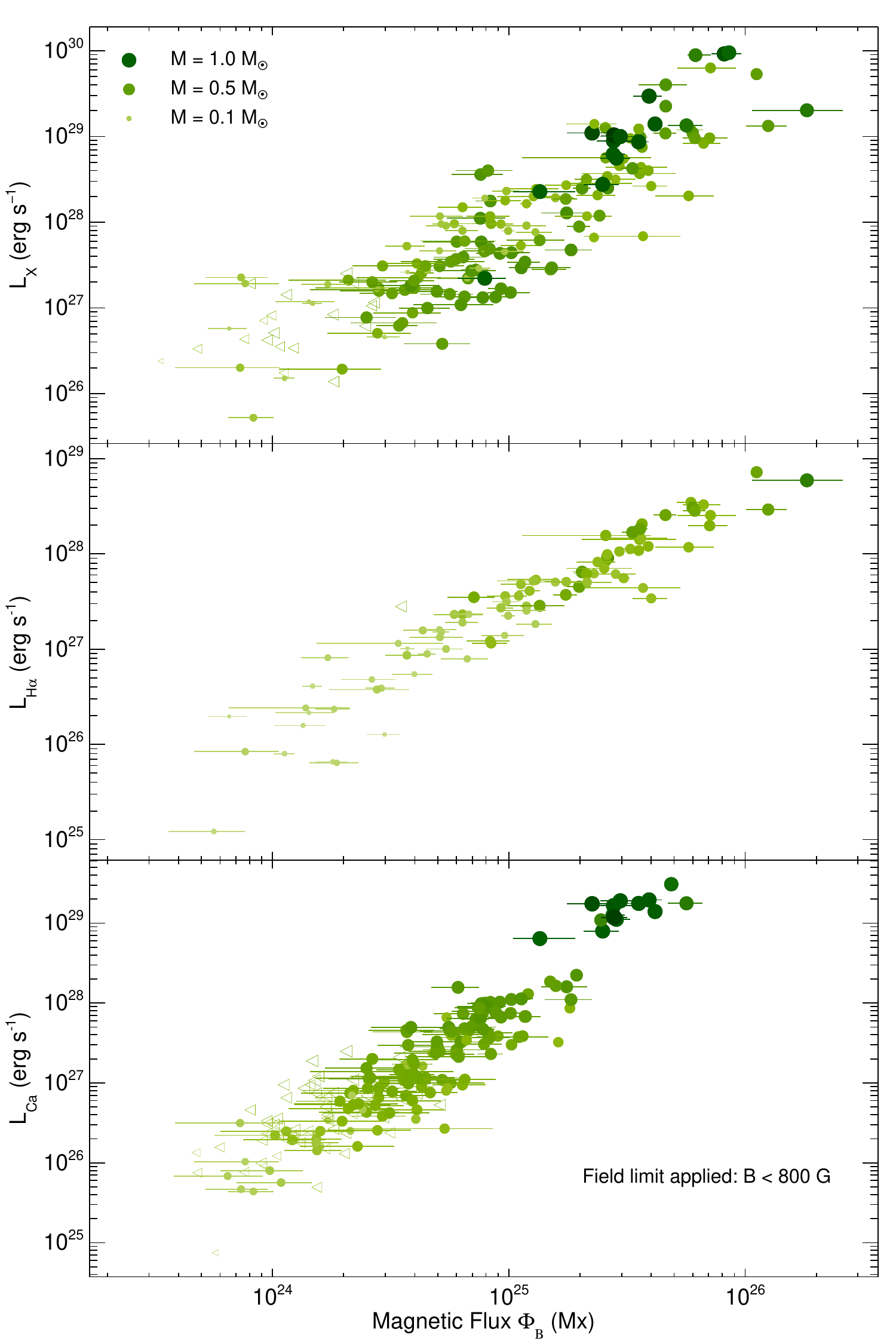}
\caption{Relations between magnetic flux, $\Phi_{\textrm B}$, and total non-thermal emission as observed in X-rays (top panel), H$\alpha$ (center panel), and Ca\,H\&K (bottom panel). Stellar mass is indicated with symbol size and color, leftward triangles indicate upper limits in $<$$B$$>$. A maximum field strength of 800\,G was used for the calculation of $\Phi_{\textrm B}$ in the bottom panel \citep[see text and discussion in][]{2022A&A...662A..41R}. \label{fig:Flux_Emission}}
\end{figure}
%%%%%%%%%%%%%%%%%%%%%%%%%%%%%%%%%%%%%%%%%%%%%%%%%%%%%%%%%%%%%%%%%%%%%%%%%%%%%%%%%%%%%%%

The magnetic field-rotation relation (Fig.\,\ref{fig:Bave_Ro}) closely
resembles the rotation-activity relation mentioned above. This suggests that
there is also a tight relation between magnetism and non-thermal emission. A
relation between X-ray luminosity and magnetic flux was reported by
\cite{2003ApJ...598.1387P} and re-investigated for a sample of stars with
measured surface fields by
\cite{2013ApJ...779..183F}. Figure\,\ref{fig:Flux_Emission} shows relations
between X-ray, H$\alpha$, and Ca\,H\&K luminosity as a function of surface
magnetic flux, $\Phi_B$. All three luminosities significantly grow with
$\Phi_B$. It should be noted that both luminosity and magnetic flux scale with radius squared, implying that at least parts of the relations seen in Fig.\,\ref{fig:Flux_Emission} \citep[and in][]{2003ApJ...598.1387P} could be caused by the different radii. An analysis of magnetic flux density and normalized luminosities reveals that higher magnetic field strengths in fact cause stronger emission \citep[see][]{2022A&A...662A..41R}.

Figures\,\ref{fig:Bave_Ro} and \ref{fig:Flux_Emission} are adapted from
\cite{2022A&A...662A..41R} and mainly cover M dwarf stars plus several young
Suns from the literature. A broad range of stellar masses and rotation rates
are included. The more rapidly rotating stars are observed to show stronger
surface fields, and stellar surface magnetic flux leads to a proportional
amount of non-thermal chromospheric and coronal emission. The Ca\,H\&K lines
show emission already at relatively low magnetic flux levels, and the field
strength adopted for the calculation of $\Phi_B$ was limited to $\langle B\rangle < 800$\,G
indicating saturation of Ca\,H\&K emission in very active stars
\citep[see][for more details]{2022A&A...662A..41R}. Coronal X-ray emission
requires a somewhat higher magnetic flux level to generate observable emission
rates, and H$\alpha$ becomes visible in emission at similar or higher magnetic
flux levels. The observed relations between magnetism and rotation, and
between magnetism and non-thermal emission provide a link between the wealth
of information on stellar rotation and stellar activity. This should help to
further constrain the processes of magnetic field generation and flux
emergence in sun-like and low-mass stars.

An important property of stellar magnetic fields is its distribution across
the stellar surface. Measurements of Zeeman broadening are sensitive to the
integrated field across the stellar surface, but they are less sensitive to
the distribution of individual field components. Measurements of polarized
light in addition with observations taken at different times, with the goal of
sampling the star at different rotational phases, can provide important
information about the geometry of the field. The field measured in polarized
light, in particular in the case of circular polarization only (Stokes~$V$),
resembles the distribution of the so-called large-scale field. This is because
magnetic fields of opposite polarity can cancel each other and remain
invisible to this type of observation. In general, results about the scaling
of large-scale fields with stellar mass and rotation are consistent with
observations of average surface fields and non-thermal emission
\citep{2014MNRAS.441.2361V}.

Observations of large-scale fields have provided insight into magnetic
geometries in very different stars \citep{2009ARA&A..47..333D,
2014MNRAS.444.3517M, 2021A&ARv..29....1K}. Additional information can be
derived from the ratio between the large-scale field from circular
polarization and the average field from Zeeman broadening. The ratio between
Stokes~V and Stokes~I field strengths is typically on the order of 10\,\% with
individual stars showing up to 40\,\% or less than 1\,\% of their magnetic fields
on large-scales. There is a slight trend of larger ratios
$\langle B\rangle_{\textrm{V}}/\langle B\rangle_{\textrm{I}}$ occuring among the lower mass / smaller
stars, which may indicate different modes of magnetic field generation but
could also be influenced by biases in the observational methods. In addition to gathering more data on magnetic fields from Stokes~$V$ and $I$ observations, it is very important to understand the systematic bias introduced by both methods. These include effects like unknown velocity fields and line profile distortions caused by non-thermal emission in Stokes~$I$ measurements, and the consequences of incomplete phase coverage, uncertainties of the inclination angle, and flux cancellation in Stokes~$V$ measurements.

\section{Modeling photospheric magnetism}\label{sec3}

%Sunspots are easy to spot, even with the smallest telescopes, thanks to the proximity of 
%the Sun. They have been monitored since Galileo, making a long time series of sunspot 
%number, area, and locations. 
Magnetic features on stars other than the Sun can only be observed 
indirectly (except in a few cases with interferometry), as mentioned in Section~\ref{sec2}. 
%Therefore, physics-based modeling of starspots is important to better evaluate our physical expectations and their confrontation with observations. 
In essence, filling factors and distributions of starspots can only be inferred from disc-
integrated diagnostics, often through data-driven modeling. Forward modeling (i.e., via 
numerical simulations) of surface magnetic features come into stage here, as they often 
bring physical insight that helps us interpret observations. This section is devoted to recent attempts in numerical simulations of observational diagnostics, based on physical models of magnetic flux generation and transport. 

\subsection{Distribution of surface magnetic flux}
\label{ssec:distro}

Attempts to model surface magnetic activity patterns on cool stars range from 
star-in-a-box simulations of convectively driven dynamos in M stars 
\citep[e.g.,][]{yadav15} to solar-like models applied to stars rotating faster than 
the Sun, which we will focus on here. 

The existence of polar or high-latitude spots on young solar-type stars rotating much 
faster than the Sun was revealed by Doppler imaging studies \citep[e.g.,][]{strassrice98}. 
The formation and structure of near-polar spots is one of the problems in cool-star 
research. The main question is whether they emerge at near-polar latitudes, or 
are transported there by surface flows. 
There are two non-mutually exclusive explanations for such spot patterns in the 
literature. Firstly, any radially rising buoyant concentration of toroidal magnetic flux 
is expected to be deflected towards the rotation axis, owing to angular momentum 
conservation (or Coriolis force in the co-rotating frame), leading to poleward deflection 
that increases with the stellar rotation rate \citep[][see also Weber et al. in this 
volume]{schuessler92}. Numerical simulations of flux-tube emergence through the convection 
zone as a function of the stellar rotation rate provided further support for this 
hypothesis \citep{schuessler96}. According to the hypothesis, poleward deflection of 
rising tubes is controlled not only by the rotation rate, but also by the fractional depth 
of the convection zone (in turn, the Rossby number). A flux loop that starts from a low 
latitude and rises parallel to the rotation axis would thus have an emergence latitude 
that increases towards later spectral types as shown in simulations by \cite{granzer00}. 
The second explanation involves the transport of emerging flux by differential rotation, 
meridional flow and supergranulation \citep{schrijver01}. Surface flux transport (SFT) 
simulations for various differential rotation and stellar radius configurations by  
\cite{isik07} have shown that, mid-latitude emergence of highly tilted bipolar regions 
sustain polar spots, mainly by diffusion and meridional flow. Faster-than-solar meridional 
flow speeds were invoked by \cite{holzwarth06}, who showed formation of possible starspots 
with intermingled polarities around the rotational pole. All these studies showed various 
possibilities for the maintenance of long-lived polar spots by SFT processes. 

Stellar dynamo models were incorporated to explain/simulate stellar cycle characteristics in addition to surface distributions. In a model that integrates a deep-seated dynamo, flux-tube rise and surface transport, \citet{isik11} estimated the evolving surface distribution of large-scale radial magnetic flux through several dynamo cycles. They found that for intermediate rotators ($P_{\rm rot}\sim 10$~d), the combined effects of enhanced cycle overlap and large tilt angles of emerging bipoles can lead to an unsigned-flux balance between polar caps and low latitudes that are modulated in anti-phase. This was suggested as an explanation for the existence of moderately active but non-cycling stars that were observed in S-index time series \citep{hall+lockwood04}. For more extensive reviews of 
modeling work on stellar activity cycles, see \citet{Biswas+23} and \citet{Hazra+23}. 

Aiming to forward-model brightness variability in rotational time scales, 
\cite{isik18} constructed a Flux Emergence and Transport (FEAT) simulation 
framework for solar-type stars with rotation rates and activity levels from 
the solar reference levels up to 8 times higher values. Assuming a solar-like 
latitudinal distribution of magnetic flux at the base of the convection zone, 
the authors modeled the surface emergence patterns by using simulations of 
buoyant flux tubes rising through the convection zone. The latitudes and tilt 
angles of emerging bipolar regions are used as input to a surface flux 
transport model, which calculates the time-dependent distribution of surface 
magnetic flux with a daily cadence, for a decade. The FEAT model also features 
active-region nesting at the time of emergence, with the probability that an 
active region emerges near the previous one being an adjustable parameter. 
They found that polar spots start to form between 4 and 8 times the solar 
rotation rate and activity level, by accumulation of trailing-polarity flux 
from tilted bipoles emerging at mid-latitudes. The tilt angles and thus 
the dipole contributions of emerging active regions increase in average, 
along with their variance around the mean \citep{Weber+23}. 
An enhanced nesting tendency locally increases the flux density reached in 
certain regions as well as in the formation of the spotted polar caps. 

\subsection{Line profile modeling}
\label{ssec:zdimodel}

Despite the currently inaccessible detailed structure of stellar active regions, solar 
observations indicate that the magnetic field should be mostly radial upon emergence into 
the photosphere, owing to the steep density gradient in the atmosphere. Along with Gauss's 
law, this enforces a bipolar distribution of radial magnetic field throughout an active 
region. However, Zeeman-Doppler imaging (ZDI) studies indicate that, as the activity level 
increases, some rapidly rotating, young cool stars manifest strong azimuthal fields 
that can form axisymmetric bands \citep[e.g.,][]{folsom18}. The azimuthal field 
component tends to follow power laws with two different exponents on both sides of 
a stellar mass of about 0.5\,M$_\odot$, and it becomes more axisymmetric and confined to 
higher latitudes for more rapid rotators \citep{See+15}. 
Strong azimuthal fields on some stars much more active than the Sun was 
interpreted by \citet{solanki02} as the effect of differential rotation acting on 
strong fields that accumulate or emerge near the rotational poles, resulting in 
amplification of azimuthal fields at mid-latitudes. Motivated by ZDI 
results, \citet{Lehmann+17} decomposed simulated surface flux 
distributions into spherical harmonics, to obtain 
relative fractions of magnetic energy in radial, azimuthal, and meridional components 
\citep[see][for a similar analysis for the Sun]{Vidotto16}.  They found that magnetic 
multipoles beyond the quadrupole (mainly contributed by azimuthal fields) host poloidal 
and toroidal field components following a fixed ratio. Later, \citet{Lehmann+19} 
synthesised Stokes profiles from surface flux transport simulations for various 
activity levels and concluded that ZDI overestimates the relative contributions from 
axisymmetric and toroidal fields, particularly as the axial inclination decreases 
towards pole-on configurations. 

As a first application of the FEAT model described in Section~\ref{ssec:distro}, 
\citet{Senavci+21} synthesised Doppler images 
of the young solar analogue EK Draconis, using SFT snapshots from the FEAT model, and 
compared them with the Doppler images they generated from observed spectra. 
The overall latitudinal distribution of spots in the simulations were consistent with the 
observations in the case of strong differential rotation, which was previously reported for 
EK Dra. The simulations also showed that low-latitude spots in observed Doppler images can 
result from mid-latitude activity in the partially visible rotational hemisphere, owing to 
the axial inclination of $63^\circ$. This study has shown the importance of 
forward modeling of the observed signals from physics-based models, in the interpretation 
of observations.

\subsection{Brightness variability}
\label{ssec:photomodel}
Broad-band variability of cool stars on time scales ranging from the rotation 
period to the activity cycle result from photospheric manifestations of magnetic 
flux: mainly spots in active regions (ARs) and faculae resulting from AR network 
fields. Intriguingly, some solar-type stars with near-solar rotation rates and 
temperatures display unexpectedly large photometric variability amplitudes on the 
rotational time scale, in comparison to the Sun \citep{Reinhold+20}. By numerical 
simulations of light curves with different modes and degrees of active-region nesting, 
\citet{Isik+20} showed that such high-amplitude light curves can be explained by a 
moderate increase of the emergence frequency and a high degree of nesting (up to 90\%). 
The same study showed that active-longitude-type nesting can reproduce light-curve 
morphology of some stars with sinusoidal-like light curves in the same sample better 
than free nesting, where nests are allowed to form within active latitudes and a 
random longitude. 

\begin{figure}
    \centering
    \includegraphics[width=.7\columnwidth]{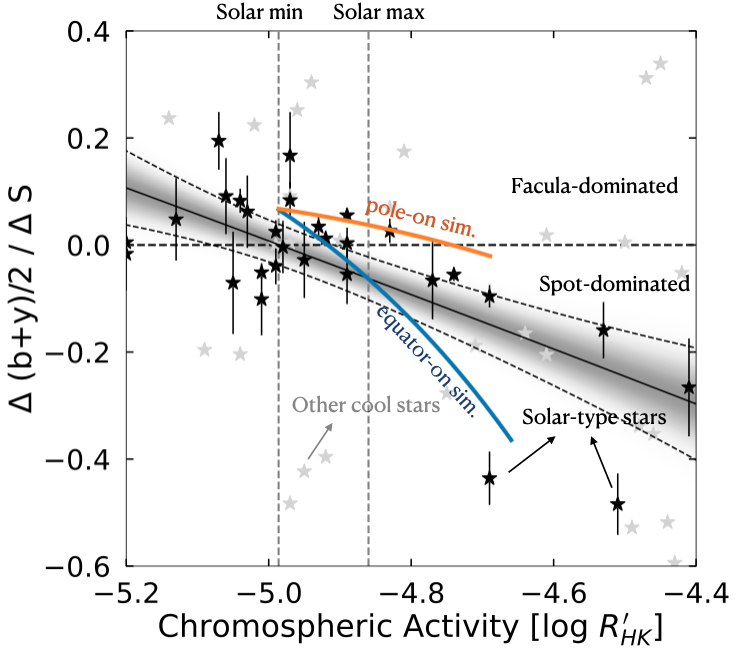}
    \caption{Activity-related brightnening (negative for dimming) as a function of the 
    mean level of chromospheric activity in the HK bands. The stellar observations 
    are shown by black stars ($T_{\rm eff}$ to within 200 K of the solar value and 
    relative brightness uncertainty below 0.01) and gray stars corresponding to other 
    stars in the sample of \citet{Radick+18}. The gray-shaded region shows the 
    posterior distribution to $2\sigma$ of Bayesian linear regression to the near-solar 
    sample, using Gaussian priors for a quadratic function. Blue and orange curves show 
    the calculated brightening functions using SFT simulations, for equator-on and 
    pole-on views, respectively.}
    \label{fig:facspot}
\end{figure}

The Sun's brightness variability on the cycle timescale is dominated by bright faculae, 
because they cover larger areas despite their visibility being confined to near-limb 
regions in visible wavelengths. However, more active stars are known to get dimmer as 
they reach peak activity levels in their S-index time series \citep[e.g.,][]{Radick+18}. 
On the Sun, the area fractions of spots and faculae depend quadratically and linearly on 
the S-index, respectively. These dependencies can be used to predict that in more active 
stars, spots would dominate over faculae in terms of disc-area coverages and hence of 
brightness \citep{Shapiro+14}. The physical mechanism underlying this transition from 
facula-dominated to spot-dominated variability has been explained by \citet{Nemec+22}, 
who carried out surface flux transport (SFT) simulations of a single solar-like activity 
cycle with gradually higher activity levels. The simulations were carried out for  
equator-on and pole-on inclinations. They showed qualitative agreement with the 
observed transition between facula- and spot-dominated cyclic variations of the 
solar-type sub-sample of \citet{Radick+18}. 
Figure~\ref{fig:facspot} shows the Str\"omgren $b+y$ brightening in units of the change 
(increase) in S-index as a function of $\log R^\prime_{\rm HK}$ for the stellar sample. 
Also shown are brightening estimates from a series of SFT simulations for two extreme 
axial inclinations. The expected transition from spot- to 
facula-dominated regime is not far from the current solar activity level during maxima. 
Here, the suggested mechanism for the transition is that for more active stars with 
presumably higher flux emergence rates, the active-region network field spanning 
much larger areas than sunspots finds more chance to undergo flux cancellation, owing to 
random superpositions of opposite magnetic polarities. On the Sun, the activity level 
is low enough to avoid such an overall flux cancellation effect, and the facular state 
remains the dominant counterpart, leading to a slight brightening of the Sun with 
increasing activity. 

Physics-based forward modeling of light curves can help constrain the parameter space of 
stellar surface brightness distributions and the axial inclination. In such an attempt, 
the FEAT model (Section~\ref{ssec:distro}) was used to calculate synthetic light curves of 
G2V stars with rotation rates between $\Omega_\odot$ and $8\Omega_\odot$ \citep{Nemec+23}. 
The method involved integration of facular and spot disc coverages weighted by the 
\emph{Kepler} transmission function, evaluated at various axial inclinations. The results 
reproduced several observed characteristics of \emph{Kepler} light curves of 
stars with different activity levels and variability patterns. However, for a better 
match to observations and for an improved understanding of the observed change in stellar 
variability patterns as a function of the rotation rate, empirical relationships
involving the magnetic flux emergence rate should be established and incorporated into 
models of flux emergence and transport. 

\subsection{Astrometric and radial velocity jitter}
\label{ssec:astrometry}
One promising method to infer activity-pattern characteristics is to make use 
of time-resolved high-precision astrometric measurements, exploiting the spatial symmetry 
breaking around the line of sight through the disc centre \citep{Lanza+08}. This 
lack of axial symmetry results from the fact that different portions of the disc have 
varying surface brightness. Based on a method developed by \citet{Shapiro+21}, numerical 
simulations of activity-induced astrometric jitter of solar-type stars were carried out 
by \citet{Sowmya2021} as a function of axial inclination, metallicity, and active-region 
nesting, finding that activity cycles can be inferred from systematic changes in the 
photocenter positions \citep[see also][]{Meunier+20}. Moreover, when the degree of 
active-region nesting is high enough, 
the cyclic changes in the photocenter jitter can be detected by Gaia. Simulations for 
more active and rapidly rotating solar-type stars have shown that the jitter becomes 
spot-dominated and could be observed even on monthly timescales \citep{Sowmya+22}. 

The radial velocity (RV) time series derived from photospheric absorption lines 
includes information on magnetic features as they transit the visible stellar disc, 
often hampering high-precision exoplanet detection with the same method. 
As part of a physics-based modeling framework of several activity indicators as a 
function of several stellar properties \citep{Meunier+19a}, the radial velocity time 
series were also modelled by \citet{Meunier+19b}. The RV variations led by magnetic 
features and their spatial distributions on the stellar disc showed general agreement 
with observations. The main features strongly affecting RV amplitudes were found to be 
the latitude coverage of active regions, the level of activity, and axial inclination.

\section{Chromospheric and Coronal Activity}\label{sec4}

The most widely used indicator of chromospheric activity in solar-type stars is known as 
the S-index, which was devised in the late-1960s at Mount Wilson Observatory 
\citep{Wilson1968}. The S-index is a measurement of the stellar flux in the cores of the 
Ca~{\sc ii} H and K spectral lines ($N_{\rm H}, N_{\rm K}$) relative to the flux in two 
neighboring pseudo-continuum bands ($N_{\rm V}, N_{\rm R}$).
 \begin{equation}
  S = \alpha \cdot \frac{N_{\rm H} + N_{\rm K}}{N_{\rm V} + N_{\rm R}},
 \end{equation} 
where $\alpha$ is a calibration constant that can be determined for any instrument using 
observations of standard stars monitored by the Mount Wilson survey \citep{Vaughan1978}. 
Time series measurements of the S-index for any given star can reveal variability due to 
stellar rotation and magnetic cycles \citep{Jeffers+23}. However, a meaningful comparison of stars with 
different spectral types requires a small correction for the photospheric contribution to 
the H and K emission, as well as a normalization by the bolometric luminosity 
\citep{Noyes1984}. This normalized activity indicator is known as $R'_{\rm HK}$, and has 
been measured for thousands of solar-type stars \citep{BoroSaikia2018}.

%%%%%%%%%%%%%%%%%%%%%%%%%%%%%%%%%%%%%%%%%%%%%%%%%%%%%%%%%%%%%%%%%%%%%%%%%%%%%%%%%%%%%%%
\begin{figure}[t]
\centering\includegraphics[width=0.9\textwidth]{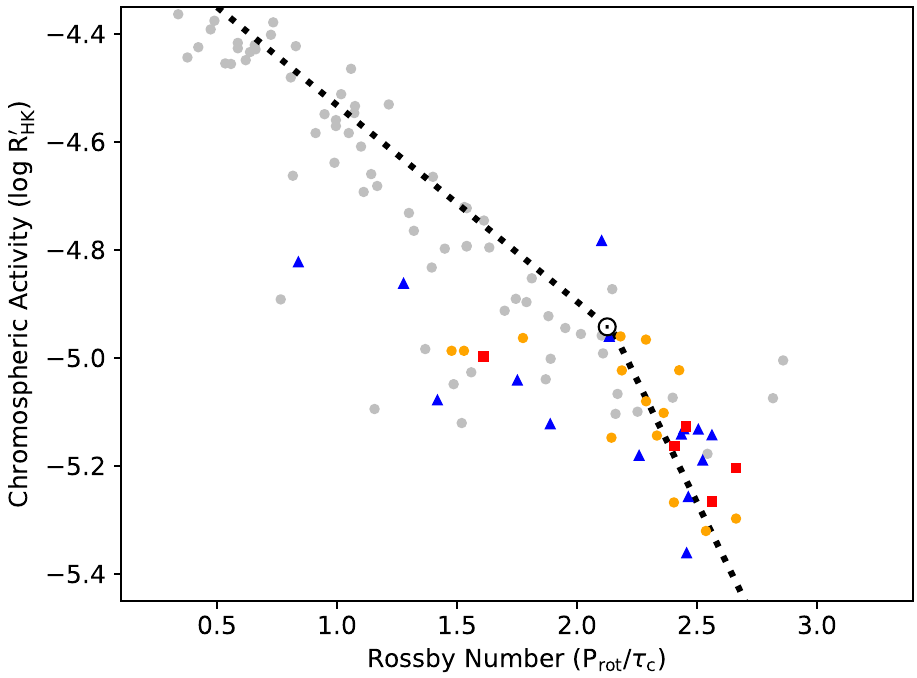}
\caption{The evolution of chromospheric activity with Rossby number for solar-type stars 
in the Mount Wilson survey (gray points) and for Kepler asteroseismic targets with 
precise Rossby numbers (colored points). Note the change in slope near the solar 
activity level.\label{fig4.1}}
\end{figure}
%%%%%%%%%%%%%%%%%%%%%%%%%%%%%%%%%%%%%%%%%%%%%%%%%%%%%%%%%%%%%%%%%%%%%%%%%%%%%%%%%%%%%%%

The evolution of $R'_{\rm HK}$ over stellar lifetimes provides one method of estimating 
ages for isolated solar-type stars \citep{Mamajek2008, Lorenzo2018}. During the first 
half of their main-sequence lifetimes, the rotation rates and activity levels of 
solar-type stars appear to decline together roughly with the square-root of the age 
\citep{Skumanich1972}. At some critical value of the Rossby number (Ro~$\equiv\!P_{\rm 
rot}/\tau_{\rm c})$ when the rotation period becomes comparable to the convective 
turnover time, rotation and activity become decoupled and the subsequent evolution of 
activity appears to be dominated by slow changes in the mechanical energy available from 
convection \citep{BohmVitense2007, Metcalfe2016}. Figure~\ref{fig4.1} shows the 
rotation-activity relation for solar-type stars in the Mount Wilson survey \citep[gray 
points;][]{Baliunas1996} and for Kepler asteroseismic targets \citep[colored 
points;][]{Metcalfe2016}. For the Mount Wilson stars, Rossby numbers have been calculated 
from $B-V$ colors \citep{Noyes1984}, and show a large scatter particularly at low 
activity levels ($\log R'_{\rm HK}\!<\!-5$). For the Kepler targets, Rossby numbers have been 
calculated using turnover times near the base of the convective envelope from an 
asteroseismic model for each star \citep{Metcalfe2014}, including hotter F-type (blue 
triangles), sun-like G-type (yellow circles), and cooler K-type stars (red squares). Most 
of this sample appears to follow a common relation at low activity levels, possibly due 
to the higher precision of their Rossby numbers. Some outliers remain at lower Rossby 
number and higher activity level, perhaps from activity variations within an unknown 
magnetic cycle. Note that the rotation-activity relation is consistent with a change in 
slope near the solar activity level, which corresponds to the apparent onset of weakened
magnetic braking \citep{vanSaders2016}.

Modeling the chromospheric emission due to magnetic activity is important to better 
understand the physical characteristics of magnetism as a function of stellar properties. 
Such models would also be useful in disentangling various mechanisms that are likely 
responsible for generating the observed distributions of chromospheric emissions. 
Recently, \citet{Sowmya+21} developed a physics-based approach to forward-model 
S-index variations of the Sun as a star, from rotational to century-scale variations, 
but for the full range of the inclination of the rotation axis with respect to the 
line of sight. Based on their calculated S-index time series of the Sun for cycles 1-23, 
they found that the variability amplitude in the chromospheric emission of 
the Sun as seen from the full range of inclinations was representative of the variability 
distribution of other solar-type stars with similar 
$\langle R^\prime_{\rm HK}\rangle$, adapted from \citet{Radick+18}. 

While magnetic heating of the chromosphere is evident in optical and ultraviolet spectral 
lines, similar processes heat the stellar corona to $\sim$1~million~K, emitting at X-ray 
wavelengths. A measurement of the X-ray luminosity is the coronal equivalent of the 
chromospheric S-index, and normalizing by the bolometric luminosity facilitates the 
comparison of stars with different spectral types like $R'_{\rm HK}$. Depending on their 
evolutionary state, solar-type stars typically have a fractional X-ray luminosity $R_{\rm 
x}\!\equiv\!L_{\rm x}/L_{\rm bol}\!\sim\!10^{-3}$ to $10^{-8}$ \citep{SchmittLiefke2004}. 
Rather than decrease monotonically with rotation rate or Rossby number, there appear to 
be different regimes in the rotation-activity relation for X-rays. For the youngest and 
most rapidly rotating stars, there is a ``saturated'' regime in which $R_{\rm x}$ appears 
relatively constant for Ro\,$<\!0.1$ \citep{Pizzolato2003}. It was initially unclear whether 
this saturation represented active regions filling the entire stellar surface, or a 
saturation of the underlying dynamo mechanism \citep{Vilhu1984}, but recent evidence 
suggests that it is a feature of the dynamo \citep{2022A&A...662A..41R}. For stars with Rossby 
numbers between $\sim$0.1 and the solar value, there is an ``unsaturated'' regime 
in which $R_{\rm x}$ decreases linearly with $\log$~Ro, following a power law with an 
index $\beta\!\sim\!2$ \citep{Wright2011}.

%%%%%%%%%%%%%%%%%%%%%%%%%%%%%%%%%%%%%%%%%%%%%%%%%%%%%%%%%%%%%%%%%%%%%%%%%%%%%%%%%%%%%%%
\begin{figure}[t]
\centering\includegraphics[width=0.9\textwidth]{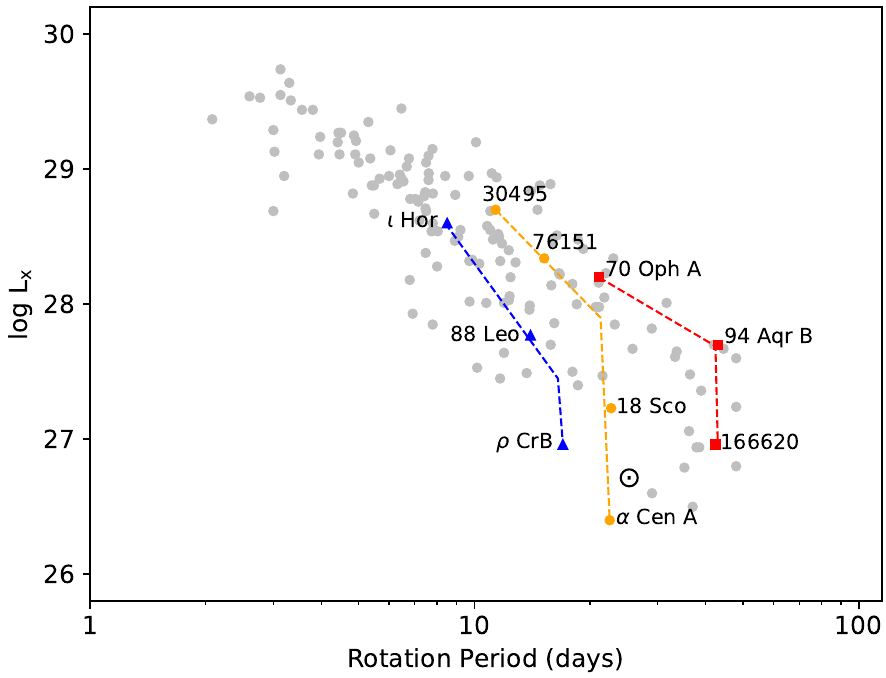}
\caption{The evolution of coronal activity with rotation for solar-type stars in the 
unsaturated regime \citep[gray points;][]{Wright2011}, with evolutionary sequences of 
F-type (blue), G-type (yellow), and K-type stars (red) for comparison. The vertical 
tracks for older stars suggest a previously unrecognized ``decoupled'' regime in the 
rotation-activity relation.\label{fig4.2}}
\end{figure}
%%%%%%%%%%%%%%%%%%%%%%%%%%%%%%%%%%%%%%%%%%%%%%%%%%%%%%%%%%%%%%%%%%%%%%%%%%%%%%%%%%%%%%%

The evolution of X-ray luminosity for Rossby numbers greater than the solar value has 
been largely unexplored, with only a few measurements available at $R_{\rm x}\!<\!10^{-6}$. 
Considering the rotational and magnetic transitions that have recently been identified 
near the solar Rossby number \citep{vanSaders2016, Metcalfe2016}, we can evaluate the 
currently available data from a new perspective. Rotation periods and X-ray luminosities 
are shown in Figure~\ref{fig4.2} for solar-type stars in the unsaturated regime 
\citep[gray points;][]{Wright2011}, with evolutionary sequences of hotter F-type (blue 
triangles), sun-like G-type (yellow circles), and cooler K-type stars (red squares) for 
comparison. When stars reach the critical rotation period that corresponds to the onset 
of weakened magnetic braking for a given spectral type, the X-ray luminosity continues to 
decline at roughly constant rotation period. For example, despite having very similar 
rotation periods, the X-ray luminosities of the solar analogs 18~Sco 
\citep[3.7~Gyr;][]{Li2012} and $\alpha$~Cen~A \citep[5.4~Gyr;][]{Bazot2016} differ by 
nearly an order of magnitude. This behavior suggests a previously unrecognized 
``decoupled'' regime in the rotation-activity relation, where the X-ray luminosity is no 
longer determined by rotation.

\section{Magnetic Braking}\label{sec5}

Magnetic braking is the loss of angular momentum (AM) through the interaction of stellar mass loss and magnetic fields. Stars with convective outer envelopes ($T_{\rm eff} < 6250$K) and magnetic dynamos spin down over time due to angular momentum loss from magnetized winds \citep{Kraft1967, Skumanich1972}. It represents a unique window into the magnetism of the Sun and stars because operates in a feedback loop: slower rotation leads to weaker magnetic fields, which lead to less AM loss via magnetized winds. It is therefore both a consequence and a driver of magnetic evolution.  Rotational evolution can occur over billion-year timescales, and represents one the best tests of the integrated behavior of the large-scale magnetic fields of stars over stellar lifetimes. Spin-down is most directly a probe of the strength of the large-scale dipole \citep{Weber1967, Kawaler1988}, with only minor contributions from higher-order fields under specific conditions \citep{See2019}. However, observed rotational evolution depends on many interacting ingredients---the magnetic field, details of the mass loss and wind flow, initial rotation rates, and internal angular momentum redistribution---making the interpretation of rotational evolution in the context of magnetism a subtle task. 

\subsection{Modeling Magnetic Braking}

A model of rotational evolution has three ingredients: 1) the braking law, which most directly probes the magnetic field behavior, 2) some assumption of the initial AM, and 3) a prescription for internal AM redistribution. Together, they predict rotational evolution as a function of time. 

\subsubsection{Braking laws}
The main sequence (MS) is where magnetic braking has the largest impact on rotational evolution for single stars. Over the MS lifetime, the braking appears to be a strong function of the rotation velocity--- $\frac{dJ}{dt} \sim \omega^3$ \citep{Skumanich1972}, which asymptotically forces convergence to a narrow range of rotation periods at a given time for a given stellar mass. This has the benefit of making the rotation rates of old stars insensitive to the significant (and for purposes here, uninteresting) spread in birth rotation periods. Observationally, this convergence happens first in the more massive stars and later in the low-mass stars, with open clusters showing tight, converged rotation sequences by a few hundred Myr around solar temperatures. 

The standard $\frac{dJ}{dt} \propto \omega^3$ spin down results in a period-age relation that goes roughly as $P_{rot} \propto \sqrt{t}$--- so-called ``Skumanich" spin down \citep{Skumanich1972}. Much of the literature on braking seeks to empirically constrain this period-age relation while being largely agnostic to the underlying physical mechanisms \citep{barnes2007, barnes2010, Mamajek2008, Angus2015}. To use braking as a constraint on magnetic field behavior, we turn to physically motivated magnetic braking laws that follow the basic formalism of \citep{Weber1967, Mestel1968} for angular momentum loss from magnetized stellar winds. The torque on a star is most simply written as:
\begin{equation}
    \tau = \dot{M} \Omega  \langle r_A\rangle^2,
\end{equation}
where $r_A$ is the Alfv\'{e}n radius where magnetic Alfv\'{e}n velocity and wind velocity are equivalent. The average radius defines the effective ``lever arm" of the torque, and depends both on the strength and morphology of the magnetic field  \citep{Reville2015, Garraffo2016,Finley2018}. In practice, the challenge in defining a braking law is in prescribing $\dot{M}$ and $B$ (as it enters in $r_A$). Many authors choose to fix a magnetic field morphology and make some assumption about how the magnetic field scales with stellar properties, often in the form of a Rossby scaling \citep[e.g.][]{Kawaler1988,Krishnamurthi1997,vanSaders2013,Matt2015}. Modern braking laws increasingly draw their forms from $>1$D MHD simulations of mass loss entrained in a magnetized wind \citep[e.g.][]{Matt2008}, but still fundamentally require some additional input of how $B$ scales with stellar properties, and necessarily make strong assumptions about the nature of wind launching. For mass loss, authors commonly adopt either empirical \citep[e.g.][]{Wood2005,Wood2021} or theoretical scalings \citep{Cranmer2011}, but the choice remains a significant uncertainty. 

\subsubsection{Additional Ingredients}

Stars are born with roughly two orders of magnitude of spread in their initial rotation periods \citep{Irwin2009, Herbst2002}, due both to stochasticity in the birth angular momentum itself and star-disk interactions that occur in the first few Myr of the star's life \citep{Matt2005, Shu1994, Koenigl1991}. While rotational evolution in solar mass stars becomes insensitive to choices of initial conditions within a few hundred Myrs, lower mass stars may retain sensitivity for Gyrs \citep{Gallet2015}, complicating the interpretation of their rotation periods \citep{Epstein2014, Roquette2021}. Authors generally either consider a range of initial rotation periods motivated by those observed in the youngest open clusters, or ``launch" their braking simulations from initial conditions defined by a benchmark cluster \citep[see][]{Somers2017,Epstein2014}. There is some evidence that stellar environments may alter the distribution of initial periods \citep{Coker2016, Roquette2021}.

Many braking prescriptions make the simplest (and often reasonably correct) assumption that internal AM transport is instantaneous compared to evolutionary timescales, resulting in solid body rotation. However, the efficiency of internal AM transport does affect the evolution of the surface rotation period. Literature braking laws that allow for interior AM transport generally use either 1) simple ``two-zone" models \citep{MacGregor1991} allow rotation of the core and convective envelope to evolve separately, coupled by AM transport over some characteristic timescale $\tau_{ce}$, or 2) allow for extra AM transport via a diffusion term \citep{Denissenkov2010,Somers2016, Spada2020} in interiors models. There is no accepted first-principles mechanism for interior AM transport that fully reproduces observations. Magnetically mediated transport at the tachocline has been proposed \citep{Oglethorpe2013}, as have internal gravity waves \citep{Denissenkov2008,Fuller2014,Somers2017,Cao2023}, but no conclusive mechanism has been identified.  

\subsection{Recent Modifications to Braking Laws}

Because of the uncertainties in both the braking process and the fundamental underlying stellar processes of mass loss and magnetic field generation, all modern braking laws are tuned via fitting parameters to observations. Improvements in the underlying physical models come largely from examining whether the braking prescription captures the observed mass and time dependence of the spin down. 

Observational benchmark systems must have both well-known ages and rotation periods. The most impactful class of calibrating systems to date has been the open clusters, which represent coeval stellar populations that span a range of masses at uniform composition. However, cluster calibrators have historically been confined to young ($<1$Gyr) ages and solar composition stars; old open clusters are rare and tend to be distant and challenging to study. Although the lack of calibrating sources at low masses, old ages, and non-solar compositions remains a persistent roadblock in the testing and validation of magnetic braking models, two classes of new calibrator sources have fueled recent refinements of magnetic braking laws: intermediate-age open clusters observed with space- and ground-based photometric missions, and bright field stars with precise, asteroseismically measured ages. Both classes have suggested significant alterations to the Skumanich-type spin-down that we will discuss below.

\begin{figure}
    \centering
    \includegraphics[width=.7\columnwidth]{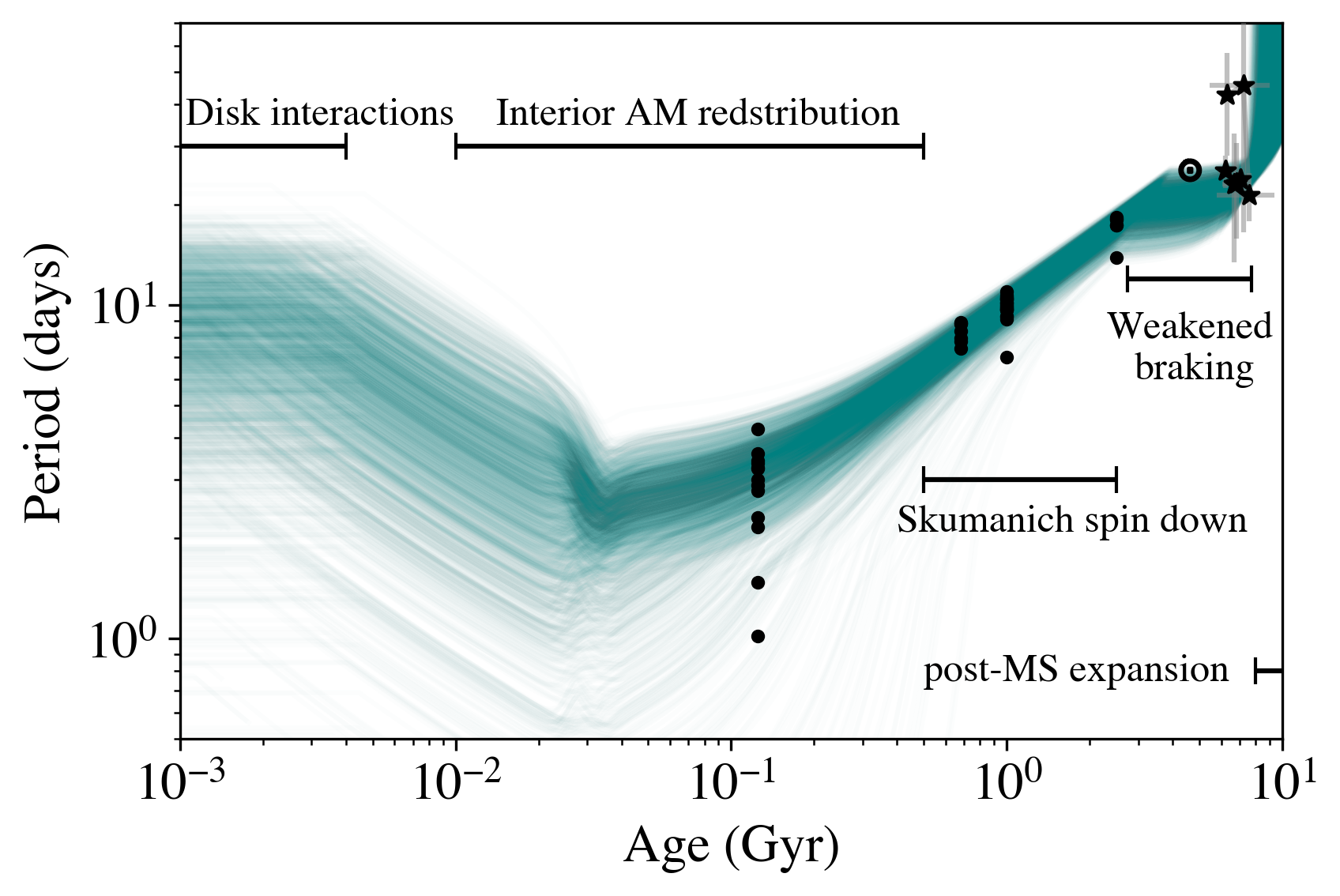}
    \caption{Calibrator stars (points) of known rotation period and age shown against calibrated braking laws that include a range of initial conditions (teal tracks), internal AM redistribution, and weakened magnetic braking. Cluster stars (circles) are drawn from the compilation in \citet{Curtis2020}, and asteroseismic targets (stars) from \citet{vanSaders2016} and \citet{Hall2021}.}
    \label{fig:sec5}
\end{figure}

We show a sample of calibrator stars selected to lie in a relatively narrow slice between $1.0 < \textrm{M}_{\odot} < 1.1$ and $0.0 < \textrm{[Fe/H]} < 0.2$ in Figure \ref{fig:sec5}. These calibrators are shown against a braking law that includes 1) a range of initial conditions (gaussian centered at 8 days, with a $1\sigma$ width of 4 days), 2) disk-locking timescales drawn uniformly from 1-5~Myr, 3) the braking law form from \citet{vanSaders2016} with the addition of core-envelope decoupling following the \citet{Somers2016} 2-zone model. Rough regions where different processes discussed in this section are important for interpreting the rotation period are noted, but readers should be aware that the exact boundaries of these regions are both mass and metallicity dependent. 

\subsubsection{Young Stars}

Young open clusters retain a subset of rapidly rotating stars for hundreds of millions of years, longer than Skumanich-type spin-down law would predict given the simultaneous existence of slow rotators. This observation motivates the inclusion of an epoch of ``saturated" spin down in nearly all magnetic braking models, in which the spin down instead goes as $\frac{dJ}{dt} \propto \omega_{crit}^{2} \omega$ \citep{Krishnamurthi1997}. This saturated regime is the analog of saturation in other magnetic proxies: the observation that beyond some $\omega_{crit}$ more rapid rotation no longer results in a stronger magnetic response, in this case reflected in the angular momentum loss. In practice, the saturation threshold is often included as a Rossby scaling, with $\textrm{Ro}_{sat} \propto (\omega_{crit} \tau_{cz})^{-1}$. A saturation threshold of $\textrm{Ro}_{sat}\sim 0.1$ is broadly consistent both with the saturation threshold in other magnetic proxies, and with the observed rotational behavior in clusters. 

\subsubsection{Intermediate-Age Stars}

In detailed cluster observations, spin-down rates depart from those expected for unsaturated, Skumanich-type spin-down in young and intermediate-age stars. Stars initially spin-down faster-than-expected \citep{Denissenkov2010}, and then ``stalled out" at intermediate ages with what appears to be minimal evolution in the surface rotation rate \citep{2018ApJ...862...33A,Curtis2020}. Although we cannot entirely rule out changes to the magnetic braking itself as the cause, internal AM transport is emerging as a leading explanation. While the epoch of ``core-envelope" decoupling is brief and subtle for solar mass stars \citep{Denissenkov2010, Gallet2015}, the apparently long timescales for transport in low-mass stars produces a stark pileup in the cluster sequences of ~1 Gyr old open clusters \citep{Curtis2020}. There are also puzzling gaps in the distribution of cool field stars at periods just long-ward of feature, which may suggest a similar phenomenon in the population at large \citep{Lu2022}. \citet{Cao2023} identified a magnetic counterpart to the observed rotational stalling, and showed that stars putatively undergoing core-envelope recoupling in the Praesepe open cluster had starkly elevated surface spot-filling fractions, which the authors argued is the consequence of the radial shears present during this epoch of AM redistribution. At a minimum, the spin-down evolution of intermediate-age stars is more complex than Skumanich; the exact mechanism (and magnetic involvement) remain areas of active inquiry.

\subsubsection{Old Stars}

Recent observations of bright field stars with precisely measured asteroseismic ages provide insight into rotation at a wider range of ages and compositions in solar mass stars. The seismic sample also displays non-Skumanich behavior: it traces the standard magnetic braking patterns at young and intermediate ages, but appears to undergo dramatically reduced angular momentum loss in the latter half of the main sequence lifetime \citep{vanSaders2016}. Followup work showed that the observed long-period edge in the field star rotation distribution \citep{McQuillan2014,Matt2015} was also consistent with weakened braking \citep{vanSaders2019,David2022}, and that the weakened braking was apparent even when rotation rates were measured methods other than spot modulation \citep{Hall2021,Masuda2022}. These braking models allow for standard evolution until some critical Rossby number is reached--- $\textrm{Ro}_{crit}$--- after which the braking is severely reduced or AM loss truncated entirely. The mechanism again remains uncertain: \citet{vanSaders2016} and \citet{Metcalfe2016} suggested that an overall weakening of the magnetic field strength and shift to higher-order morphologies could be responsible, and detailed studies of stars on either side of the transition have thus far supported the picture of weakening dipole fields as a driver for the change in spin-down \citep[see][and references therein]{Metcalfe2023}. However, changes in the mass loss rates, and details of the wind launching and flow are not ruled out.

\section{Outlook}\label{sec6}

Because the underlying mechanisms are still poorly understood, progress in understanding stellar magnetism is very much driven by observations, and then theoretical efforts to reproduce the patterns we see. Larger, more complete, and more comprehensive datasets drive this progress. Those data are challenging to obtain---whether they be subtle spot signatures, direct measurements of weak fields, precise ages and rotation rates, or decades-long activity cycles---with significant progress in the last decade. There are two classes of observational benchmarks that enable progress: small, but exquisitely studied samples of stars with multiple precision measurements of magnetic proxies that allow us to build a complete picture of their behavior, and truly large but comparatively more poorly constrained samples that allow us to probe the properties and patterns in populations. 

The various attempts at forward modeling of photospheric activity diagnostics 
are helpful in deepening our understanding of the physics of magnetic activity, 
along with qualitative comparisons with the available observations. 
Extending magnetic flux emergence and subsequent surface evolution models 
of spots and faculae to wide ranges of stellar properties (e.g., $T_{\rm eff}$, 
differential rotation, metallicity) will be important in evaluating observational 
data. 
We note, 
however, that care should be taken when interpreting observations with the matching 
simulations. The inverse problem of recovering the surface magnetic patterns 
is often ill-posed, involving several parameter degeneracies. Likelihood-free 
inference frameworks have the potential to illuminate parameter ranges that optimally 
simulate observed data, allowing a Bayesian way of inverting observations 
\citep[e.g.,][]{Cranmer20}. 

Models of magnetic braking have seen frequent, large revisions in recent years as new data become available. Because the number of calibrating sources still spans a relatively narrow range of masses, ages, and compositions, there is likely much to learn as new corners of parameter space become accessible. Observational improvements are likely to come from a combination of many datasets, with two paths for growth: 1) small samples with exceptional observational coverage (rotation, asteroseismology, magnetic activity cycle measurements, direct magnetic field mapping for all targets), and 2) very large but more poorly characterized stellar samples, enabled by large photometric, spectroscopic, and astrometric surveys. Pushing to lower masses, older ages, and less solar-like compositions are the most critical directions for new observational insight into braking behavior. On the theoretical side, progress can be made by increasingly ground braking models in first-principles or semi-empirical models of magnetized winds, and gradually phasing out the more purely empirical scalings currently in use \citep[e.g.,][]{Chebly+23,Evensberget+23}. As has been the case since the beginning of this field, the interplay between improved observational calibrator sets and more sophisticated physical models will drive progress.

\backmatter
\bmhead{Acknowledgments}
This review was written following the workshop ``Solar
and Stellar Dynamos: A New Era'', hosted and supported by the International
Space Science Institute (ISSI) in Bern, Switzerland. The authors wish to express
their thanks to ISSI for their financial and logistical support.\\

\section*{Ethics Declarations}
{\bf Competing interests}
\medskip
The authors declare they have no conflicts of interest.

\bibliography{main_activity}

\begin{thebibliography}{131}
\providecommand{\natexlab}[1]{#1}
\providecommand{\url}[1]{{#1}}
\providecommand{\urlprefix}{URL }
\providecommand{\doi}[1]{\url{https://doi.org/#1}}
\providecommand{\eprint}[2][]{\url{#2}}
 \bibcommenthead

\bibitem[{{Ag{\"u}eros} et~al(2018){Ag{\"u}eros}, {Bowsher}, {Bochanski},
  {Cargile}, {Covey}, {Douglas}, {Kraus}, {Kundert}, {Law}, {Ahmadi}, and
  {Arce}}]{2018ApJ...862...33A}
{Ag{\"u}eros} MA, {Bowsher} EC, {Bochanski} JJ, et~al (2018) {A New Look at an
  Old Cluster: The Membership, Rotation, and Magnetic Activity of Low-mass
  Stars in the 1.3 Gyr Old Open Cluster NGC 752}. \apj 862(1):33.
  \doi{10.3847/1538-4357/aac6ed},
  {\href{https://arxiv.org/abs/1804.02016}{{https://arxiv.org/abs/arXiv:1804.02016}}}
  {[astro-ph.SR]}

\bibitem[{{Angus} et~al(2015){Angus}, {Aigrain}, {Foreman-Mackey}, and
  {McQuillan}}]{Angus2015}
{Angus} R, {Aigrain} S, {Foreman-Mackey} D, et~al (2015) {Calibrating
  gyrochronology using Kepler asteroseismic targets}. \mnras 450(2):1787--1798.
  \doi{10.1093/mnras/stv423},
  {\href{https://arxiv.org/abs/1502.06965}{{https://arxiv.org/abs/arXiv:1502.06965}}}
  {[astro-ph.EP]}

\bibitem[{{Auri{\`e}re}(2003)}]{Auriere03}
{Auri{\`e}re} M (2003) {Stellar Polarimetry with NARVAL}. In: {Arnaud} J,
  {Meunier} N (eds) EAS Publications Series, p 105

\bibitem[{{Baliunas} et~al(1996){Baliunas}, {Sokoloff}, and
  {Soon}}]{Baliunas1996}
{Baliunas} S, {Sokoloff} D, {Soon} W (1996) {Magnetic Field and Rotation in
  Lower Main-Sequence Stars: an Empirical Time-dependent Magnetic Bode's
  Relation?} \apjl 457:L99. \doi{10.1086/309891}

\bibitem[{{Barnes}(2007)}]{barnes2007}
{Barnes} SA (2007) {Ages for Illustrative Field Stars Using Gyrochronology:
  Viability, Limitations, and Errors}. \apj 669(2):1167--1189.
  \doi{10.1086/519295},
  {\href{https://arxiv.org/abs/0704.3068}{{https://arxiv.org/abs/arXiv:0704.3068}}}
  {[astro-ph]}

\bibitem[{{Barnes}(2010)}]{barnes2010}
{Barnes} SA (2010) {A Simple Nonlinear Model for the Rotation of Main-sequence
  Cool Stars. I. Introduction, Implications for Gyrochronology, and
  Color-Period Diagrams}. \apj 722(1):222--234.
  \doi{10.1088/0004-637X/722/1/222}

\bibitem[{{Basri}(2021)}]{Basri21}
{Basri} G (2021) {An Introduction to Stellar Magnetic Activity}. {IOP
  Publishing, Bristol}, \doi{10.1088/2514-3433/ac2956}

\bibitem[{{Bazot} et~al(2016){Bazot}, {Christensen-Dalsgaard}, {Gizon}, and
  {Benomar}}]{Bazot2016}
{Bazot} M, {Christensen-Dalsgaard} J, {Gizon} L, et~al (2016) {On the uncertain
  nature of the core of {\ensuremath{\alpha}} Cen A}. \mnras 460(2):1254--1269.
  \doi{10.1093/mnras/stw921},
  {\href{https://arxiv.org/abs/1603.07583}{{https://arxiv.org/abs/arXiv:1603.07583}}}
  {[astro-ph.SR]}

\bibitem[{{Berdyugina}(2005)}]{2005LRSP....2....8B}
{Berdyugina} SV (2005) {Starspots: A Key to the Stellar Dynamo}. Living Reviews
  in Solar Physics 2(1):8. \doi{10.12942/lrsp-2005-8}

\bibitem[{{Biswas} et~al(2023){Biswas}, {Karak}, {Usoskin}, and
  {Weisshaar}}]{Biswas+23}
{Biswas} A, {Karak} BB, {Usoskin} I, et~al (2023) {Long-Term Modulation of
  Solar Cycles}. \ssr 219(3):19. \doi{10.1007/s11214-023-00968-w},
  {\href{https://arxiv.org/abs/2302.14845}{{https://arxiv.org/abs/arXiv:2302.14845}}}
  {[astro-ph.SR]}

\bibitem[{{B{\"o}hm-Vitense}(2007)}]{BohmVitense2007}
{B{\"o}hm-Vitense} E (2007) {Chromospheric Activity in G and K Main-Sequence
  Stars, and What It Tells Us about Stellar Dynamos}. \apj 657(1):486--493.
  \doi{10.1086/510482}

\bibitem[{{Boro Saikia} et~al(2018){Boro Saikia}, {Marvin}, {Jeffers},
  {Reiners}, {Cameron}, {Marsden}, {Petit}, {Warnecke}, and
  {Yadav}}]{BoroSaikia2018}
{Boro Saikia} S, {Marvin} CJ, {Jeffers} SV, et~al (2018) {Chromospheric
  activity catalogue of 4454 cool stars. Questioning the active branch of
  stellar activity cycles}. \aap 616:A108. \doi{10.1051/0004-6361/201629518},
  {\href{https://arxiv.org/abs/1803.11123}{{https://arxiv.org/abs/arXiv:1803.11123}}}
  {[astro-ph.SR]}

\bibitem[{{Cao} et~al(2023){Cao}, {Pinsonneault}, and {van Saders}}]{Cao2023}
{Cao} L, {Pinsonneault} MH, {van Saders} JL (2023) {Core-envelope decoupling
  drives radial shear dynamos in cool stars}. arXiv e-prints arXiv:2301.07716.
  \doi{10.48550/arXiv.2301.07716},
  {\href{https://arxiv.org/abs/2301.07716}{{https://arxiv.org/abs/arXiv:2301.07716}}}
  {[astro-ph.SR]}

\bibitem[{{Chebly} et~al(2023){Chebly}, {Alvarado-G{\'o}mez},
  {Poppenh{\"a}ger}, and {Garraffo}}]{Chebly+23}
{Chebly} JJ, {Alvarado-G{\'o}mez} JD, {Poppenh{\"a}ger} K, et~al (2023)
  {Numerical quantification of the wind properties of cool main sequence
  stars}. \mnras 524(4):5060--5079. \doi{10.1093/mnras/stad2100},
  {\href{https://arxiv.org/abs/2307.04615}{{https://arxiv.org/abs/arXiv:2307.04615}}}
  {[astro-ph.SR]}

\bibitem[{{Coker} et~al(2016){Coker}, {Pinsonneault}, and
  {Terndrup}}]{Coker2016}
{Coker} CT, {Pinsonneault} M, {Terndrup} DM (2016) {Evidence for Cluster to
  Cluster Variations in Low-mass Stellar Rotational Evolution}. \apj
  833(1):122. \doi{10.3847/1538-4357/833/1/122},
  {\href{https://arxiv.org/abs/1604.05729}{{https://arxiv.org/abs/arXiv:1604.05729}}}
  {[astro-ph.SR]}

\bibitem[{{Cranmer} et~al(2020){Cranmer}, {Brehmer}, and {Louppe}}]{Cranmer20}
{Cranmer} K, {Brehmer} J, {Louppe} G (2020) {The frontier of simulation-based
  inference}. Proceedings of the National Academy of Science
  117(48):30,055--30,062. \doi{10.1073/pnas.1912789117},
  {\href{https://arxiv.org/abs/1911.01429}{{https://arxiv.org/abs/arXiv:1911.01429}}}
  {[stat.ML]}

\bibitem[{{Cranmer} and {Saar}(2011)}]{Cranmer2011}
{Cranmer} SR, {Saar} SH (2011) {Testing a Predictive Theoretical Model for the
  Mass Loss Rates of Cool Stars}. \apj 741(1):54.
  \doi{10.1088/0004-637X/741/1/54},
  {\href{https://arxiv.org/abs/1108.4369}{{https://arxiv.org/abs/arXiv:1108.4369}}}
  {[astro-ph.SR]}

\bibitem[{{Curtis} et~al(2020){Curtis}, {Ag{\"u}eros}, {Matt}, {Covey},
  {Douglas}, {Angus}, {Saar}, {Cody}, {Vanderburg}, {Law}, {Kraus}, {Latham},
  {Baranec}, {Riddle}, {Ziegler}, {Lund}, {Torres}, {Meibom}, {Aguirre}, and
  {Wright}}]{Curtis2020}
{Curtis} JL, {Ag{\"u}eros} MA, {Matt} SP, et~al (2020) {When Do Stalled Stars
  Resume Spinning Down? Advancing Gyrochronology with Ruprecht 147}. \apj
  904(2):140. \doi{10.3847/1538-4357/abbf58},
  {\href{https://arxiv.org/abs/2010.02272}{{https://arxiv.org/abs/arXiv:2010.02272}}}
  {[astro-ph.SR]}

\bibitem[{{David} et~al(2022){David}, {Angus}, {Curtis}, {van Saders},
  {Colman}, {Contardo}, {Lu}, and {Zinn}}]{David2022}
{David} TJ, {Angus} R, {Curtis} JL, et~al (2022) {Further Evidence of Modified
  Spin-down in Sun-like Stars: Pileups in the Temperature-Period Distribution}.
  \apj 933(1):114. \doi{10.3847/1538-4357/ac6dd3},
  {\href{https://arxiv.org/abs/2203.08920}{{https://arxiv.org/abs/arXiv:2203.08920}}}
  {[astro-ph.SR]}

\bibitem[{{Denissenkov} et~al(2008){Denissenkov}, {Pinsonneault}, and
  {MacGregor}}]{Denissenkov2008}
{Denissenkov} PA, {Pinsonneault} M, {MacGregor} KB (2008) {What Prevents
  Internal Gravity Waves from Disturbing the Solar Uniform Rotation?} \apj
  684(1):757--769. \doi{10.1086/589502},
  {\href{https://arxiv.org/abs/0801.3622}{{https://arxiv.org/abs/arXiv:0801.3622}}}
  {[astro-ph]}

\bibitem[{{Denissenkov} et~al(2010){Denissenkov}, {Pinsonneault}, {Terndrup},
  and {Newsham}}]{Denissenkov2010}
{Denissenkov} PA, {Pinsonneault} M, {Terndrup} DM, et~al (2010) {Angular
  Momentum Transport in Solar-type Stars: Testing the Timescale for
  Core-Envelope Coupling}. \apj 716(2):1269--1287.
  \doi{10.1088/0004-637X/716/2/1269},
  {\href{https://arxiv.org/abs/0911.1121}{{https://arxiv.org/abs/arXiv:0911.1121}}}
  {[astro-ph.SR]}

\bibitem[{{Donati} and {Landstreet}(2009)}]{2009ARA&A..47..333D}
{Donati} JF, {Landstreet} JD (2009) {Magnetic Fields of Nondegenerate Stars}.
  \araa 47:333--370. \doi{10.1146/annurev-astro-082708-101833},
  {\href{https://arxiv.org/abs/0904.1938}{{https://arxiv.org/abs/arXiv:0904.1938}}}
  {[astro-ph.SR]}

\bibitem[{{Donati} et~al(2006){Donati}, {Catala}, {Landstreet}, and
  {Petit}}]{Donati+06}
{Donati} JF, {Catala} C, {Landstreet} JD, et~al (2006) {ESPaDOnS: The New
  Generation Stellar Spectro-Polarimeter. Performances and First Results}. In:
  {Casini} R, {Lites} BW (eds) Solar Polarization 4, p 362

\bibitem[{{Donati} et~al(2020){Donati}, {Kouach}, {Moutou}, {Doyon},
  {Delfosse}, {Artigau}, {Baratchart}, {Lacombe}, {Barrick}, {H{\'e}brard},
  {Bouchy}, {Saddlemyer}, {Par{\`e}s}, {Rabou}, {Micheau}, {Dolon}, {Reshetov},
  {Challita}, {Carmona}, {Striebig}, {Thibault}, {Martioli}, {Cook},
  {Fouqu{\'e}}, {Vermeulen}, {Wang}, {Arnold}, {Pepe}, {Boisse}, {Figueira},
  {Bouvier}, {Ray}, {Feugeade}, {Morin}, {Alencar}, {Hobson}, {Castilho},
  {Udry}, {Santos}, {Hernandez}, {Benedict}, {Vall{\'e}e}, {Gallou}, {Dupieux},
  {Larrieu}, {Perruchot}, {Sottile}, {Moreau}, {Usher}, {Baril}, {Wildi},
  {Chazelas}, {Malo}, {Bonfils}, {Loop}, {Kerley}, {Wevers}, {Dunn}, {Pazder},
  {Macdonald}, {Dubois}, {Carri{\'e}}, {Valentin}, {Henault}, {Yan}, and
  {Steinmetz}}]{Donati20}
{Donati} JF, {Kouach} D, {Moutou} C, et~al (2020) {SPIRou: NIR velocimetry and
  spectropolarimetry at the CFHT}. \mnras 498(4):5684--5703.
  \doi{10.1093/mnras/staa2569},
  {\href{https://arxiv.org/abs/2008.08949}{{https://arxiv.org/abs/arXiv:2008.08949}}}
  {[astro-ph.IM]}

\bibitem[{{Dorn} et~al(2014){Dorn}, {Anglada-Escude}, {Baade}, {Bristow},
  {Follert}, {Gojak}, {Grunhut}, {Hatzes}, {Heiter}, {Hilker}, {Ives}, {Jung},
  {K{\"a}ufl}, {Kerber}, {Klein}, {Lizon}, {Lockhart}, {L{\"o}winger},
  {Marquart}, {Oliva}, {Origlia}, {Pasquini}, {Paufique}, {Piskunov}, {Pozna},
  {Reiners}, {Smette}, {Smoker}, {Seemann}, {Stempels}, and
  {Valenti}}]{2014Msngr.156....7D}
{Dorn} RJ, {Anglada-Escude} G, {Baade} D, et~al (2014) {CRIRES+: Exploring the
  Cold Universe at High Spectral Resolution}. The Messenger 156:7--11

\bibitem[{{Engvold} et~al(2019){Engvold}, {Vial}, and {Skumanich}}]{Engvold19}
{Engvold} O, {Vial} JC, {Skumanich} A (2019) {The Sun as a Guide to Stellar
  Physics}. {Elsevier, Amsterdam}, \doi{10.1016/C2017-0-01365-4}

\bibitem[{{Epstein} and {Pinsonneault}(2014)}]{Epstein2014}
{Epstein} CR, {Pinsonneault} MH (2014) {How Good a Clock is Rotation? The
  Stellar Rotation-Mass-Age Relationship for Old Field Stars}. \apj 780(2):159.
  \doi{10.1088/0004-637X/780/2/159},
  {\href{https://arxiv.org/abs/1203.1618}{{https://arxiv.org/abs/arXiv:1203.1618}}}
  {[astro-ph.SR]}

\bibitem[{{Evensberget} et~al(2023){Evensberget}, {Marsden}, {Carter},
  {Salmeron}, {Vidotto}, {Folsom}, {Kavanagh}, {Pineda}, {Driessen}, and
  {Strickert}}]{Evensberget+23}
{Evensberget} D, {Marsden} SC, {Carter} BD, et~al (2023) {The winds of young
  Solar-type stars in the Pleiades, AB Doradus, Columba, and
  {\ensuremath{\beta}} Pictoris}. \mnras 524(2):2042--2063.
  \doi{10.1093/mnras/stad1650},
  {\href{https://arxiv.org/abs/2305.17427}{{https://arxiv.org/abs/arXiv:2305.17427}}}
  {[astro-ph.SR]}

\bibitem[{{Feiden} and {Chaboyer}(2013)}]{2013ApJ...779..183F}
{Feiden} GA, {Chaboyer} B (2013) {Magnetic Inhibition of Convection and the
  Fundamental Properties of Low-mass Stars. I. Stars with a Radiative Core}.
  \apj 779(2):183. \doi{10.1088/0004-637X/779/2/183},
  {\href{https://arxiv.org/abs/1309.0033}{{https://arxiv.org/abs/arXiv:1309.0033}}}
  {[astro-ph.SR]}

\bibitem[{{Finley} and {Matt}(2018)}]{Finley2018}
{Finley} AJ, {Matt} SP (2018) {The Effect of Combined Magnetic Geometries on
  Thermally Driven Winds. II. Dipolar, Quadrupolar, and Octupolar Topologies}.
  \apj 854(2):78. \doi{10.3847/1538-4357/aaaab5},
  {\href{https://arxiv.org/abs/1801.07662}{{https://arxiv.org/abs/arXiv:1801.07662}}}
  {[astro-ph.SR]}

\bibitem[{{Folsom} et~al(2018){Folsom}, {Bouvier}, {Petit}, {L{\`e}bre},
  {Amard}, {Palacios}, {Morin}, {Donati}, and {Vidotto}}]{folsom18}
{Folsom} CP, {Bouvier} J, {Petit} P, et~al (2018) {The evolution of surface
  magnetic fields in young solar-type stars II: the early main sequence
  (250-650 Myr)}. \mnras 474(4):4956--4987. \doi{10.1093/mnras/stx3021},
  {\href{https://arxiv.org/abs/1711.08636}{{https://arxiv.org/abs/arXiv:1711.08636}}}
  {[astro-ph.SR]}

\bibitem[{{Fuller} et~al(2014){Fuller}, {Lecoanet}, {Cantiello}, and
  {Brown}}]{Fuller2014}
{Fuller} J, {Lecoanet} D, {Cantiello} M, et~al (2014) {Angular Momentum
  Transport via Internal Gravity Waves in Evolving Stars}. \apj 796(1):17.
  \doi{10.1088/0004-637X/796/1/17},
  {\href{https://arxiv.org/abs/1409.6835}{{https://arxiv.org/abs/arXiv:1409.6835}}}
  {[astro-ph.SR]}

\bibitem[{{Gallet} and {Bouvier}(2015)}]{Gallet2015}
{Gallet} F, {Bouvier} J (2015) {Improved angular momentum evolution model for
  solar-like stars. II. Exploring the mass dependence}. \aap 577:A98.
  \doi{10.1051/0004-6361/201525660},
  {\href{https://arxiv.org/abs/1502.05801}{{https://arxiv.org/abs/arXiv:1502.05801}}}
  {[astro-ph.SR]}

\bibitem[{{Garraffo} et~al(2016){Garraffo}, {Drake}, and
  {Cohen}}]{Garraffo2016}
{Garraffo} C, {Drake} JJ, {Cohen} O (2016) {The missing magnetic morphology
  term in stellar rotation evolution}. \aap 595:A110.
  \doi{10.1051/0004-6361/201628367},
  {\href{https://arxiv.org/abs/1607.06096}{{https://arxiv.org/abs/arXiv:1607.06096}}}
  {[astro-ph.SR]}

\bibitem[{{Granzer} et~al(2000){Granzer}, {Sch{\"u}ssler}, {Caligari}, and
  {Strassmeier}}]{granzer00}
{Granzer} T, {Sch{\"u}ssler} M, {Caligari} P, et~al (2000) {Distribution of
  starspots on cool stars. II. Pre-main-sequence and ZAMS stars between 0.4
  M$_{sun}$ and 1.7 M$_{sun}$}. \aap 355:1087--1097

\bibitem[{{Hall} and {Lockwood}(2004)}]{hall+lockwood04}
{Hall} JC, {Lockwood} GW (2004) {The Chromospheric Activity and Variability of
  Cycling and Flat Activity Solar-Analog Stars}. \apj 614(2):942--946.
  \doi{10.1086/423926}

\bibitem[{{Hall} et~al(2021){Hall}, {Davies}, {van Saders}, {Nielsen}, {Lund},
  {Chaplin}, {Garc{\'\i}a}, {Amard}, {Breimann}, {Khan}, {See}, and
  {Tayar}}]{Hall2021}
{Hall} OJ, {Davies} GR, {van Saders} J, et~al (2021) {Weakened magnetic braking
  supported by asteroseismic rotation rates of Kepler dwarfs}. Nature Astronomy
  5:707--714. \doi{10.1038/s41550-021-01335-x},
  {\href{https://arxiv.org/abs/2104.10919}{{https://arxiv.org/abs/arXiv:2104.10919}}}
  {[astro-ph.SR]}

\bibitem[{{Hazra} et~al(2023){Hazra}, {Nandy}, {Kitchatinov}, and
  {Choudhuri}}]{Hazra+23}
{Hazra} G, {Nandy} D, {Kitchatinov} L, et~al (2023) {Mean Field Models of Flux
  Transport Dynamo and Meridional Circulation in the Sun and Stars}. \ssr
  219(5):39. \doi{10.1007/s11214-023-00982-y},
  {\href{https://arxiv.org/abs/2302.09390}{{https://arxiv.org/abs/arXiv:2302.09390}}}
  {[astro-ph.SR]}

\bibitem[{{Herbst} et~al(2002){Herbst}, {Bailer-Jones}, {Mundt},
  {Meisenheimer}, and {Wackermann}}]{Herbst2002}
{Herbst} W, {Bailer-Jones} CAL, {Mundt} R, et~al (2002) {Stellar rotation and
  variability in the Orion Nebula Cluster}. \aap 396:513--532.
  \doi{10.1051/0004-6361:20021362}

\bibitem[{{Holzwarth} et~al(2006){Holzwarth}, {Mackay}, and
  {Jardine}}]{holzwarth06}
{Holzwarth} V, {Mackay} DH, {Jardine} M (2006) {The impact of meridional
  circulation on stellar butterfly diagrams and polar caps}. \mnras
  369(4):1703--1718. \doi{10.1111/j.1365-2966.2006.10407.x},
  {\href{https://arxiv.org/abs/astro-ph/0604102}{{https://arxiv.org/abs/arXiv:astro-ph/0604102}}}
  {[astro-ph]}

\bibitem[{{I{\c{s}}{\i}k} et~al(2007){I{\c{s}}{\i}k}, {Sch{\"u}ssler}, and
  {Solanki}}]{isik07}
{I{\c{s}}{\i}k} E, {Sch{\"u}ssler} M, {Solanki} SK (2007) {Magnetic flux
  transport on active cool stars and starspot lifetimes}. \aap
  464(3):1049--1057. \doi{10.1051/0004-6361:20066623},
  {\href{https://arxiv.org/abs/astro-ph/0612399}{{https://arxiv.org/abs/arXiv:astro-ph/0612399}}}
  {[astro-ph]}

\bibitem[{{I{\c{s}}{\i}k} et~al(2011){I{\c{s}}{\i}k}, {Schmitt}, and
  {Sch{\"u}ssler}}]{isik11}
{I{\c{s}}{\i}k} E, {Schmitt} D, {Sch{\"u}ssler} M (2011) {Magnetic flux
  generation and transport in cool stars}. \aap 528:A135.
  \doi{10.1051/0004-6361/201014501},
  {\href{https://arxiv.org/abs/1102.0569}{{https://arxiv.org/abs/arXiv:1102.0569}}}
  {[astro-ph.SR]}

\bibitem[{{I{\c{s}}{\i}k} et~al(2018){I{\c{s}}{\i}k}, {Solanki}, {Krivova}, and
  {Shapiro}}]{isik18}
{I{\c{s}}{\i}k} E, {Solanki} SK, {Krivova} NA, et~al (2018) {Forward modelling
  of brightness variations in Sun-like stars. I. Emergence and surface
  transport of magnetic flux}. \aap 620:A177.
  \doi{10.1051/0004-6361/201833393},
  {\href{https://arxiv.org/abs/1810.06728}{{https://arxiv.org/abs/arXiv:1810.06728}}}
  {[astro-ph.SR]}

\bibitem[{{I{\c{s}}{\i}k} et~al(2020){I{\c{s}}{\i}k}, {Shapiro}, {Solanki}, and
  {Krivova}}]{Isik+20}
{I{\c{s}}{\i}k} E, {Shapiro} AI, {Solanki} SK, et~al (2020) {Amplification of
  Brightness Variability by Active-region Nesting in Solar-like Stars}. \apjl
  901(1):L12. \doi{10.3847/2041-8213/abb409},
  {\href{https://arxiv.org/abs/2009.00692}{{https://arxiv.org/abs/arXiv:2009.00692}}}
  {[astro-ph.SR]}

\bibitem[{{Irwin} and {Bouvier}(2009)}]{Irwin2009}
{Irwin} J, {Bouvier} J (2009) {The rotational evolution of low-mass stars}. In:
  {Mamajek} EE, {Soderblom} DR, {Wyse} RFG (eds) The Ages of Stars, pp
  363--374, \doi{10.1017/S1743921309032025}, \eprint{0901.3342}

\bibitem[{{Jeffers} et~al(2023){Jeffers}, {Kiefer}, and
  {Metcalfe}}]{Jeffers+23}
{Jeffers} SV, {Kiefer} R, {Metcalfe} TS (2023) {Stellar Activity Cycles}. arXiv
  e-prints arXiv:2309.14138. \doi{10.48550/arXiv.2309.14138},
  {\href{https://arxiv.org/abs/2309.14138}{{https://arxiv.org/abs/arXiv:2309.14138}}}
  {[astro-ph.SR]}

\bibitem[{{Kawaler}(1988)}]{Kawaler1988}
{Kawaler} SD (1988) {Angular Momentum Loss in Low-Mass Stars}. \apj 333:236.
  \doi{10.1086/166740}

\bibitem[{{Kochukhov}(2021)}]{2021A&ARv..29....1K}
{Kochukhov} O (2021) {Magnetic fields of M dwarfs}. \aapr 29(1):1.
  \doi{10.1007/s00159-020-00130-3},
  {\href{https://arxiv.org/abs/2011.01781}{{https://arxiv.org/abs/arXiv:2011.01781}}}
  {[astro-ph.SR]}

\bibitem[{{Kochukhov} and {Reiners}(2020)}]{2020ApJ...902...43K}
{Kochukhov} O, {Reiners} A (2020) {The Magnetic Field of the Active
  Planet-hosting M Dwarf AU Mic}. \apj 902(1):43.
  \doi{10.3847/1538-4357/abb2a2},
  {\href{https://arxiv.org/abs/2008.10668}{{https://arxiv.org/abs/arXiv:2008.10668}}}
  {[astro-ph.SR]}

\bibitem[{{Koenigl}(1991)}]{Koenigl1991}
{Koenigl} A (1991) {Disk Accretion onto Magnetic T Tauri Stars}. \apjl 370:L39.
  \doi{10.1086/185972}

\bibitem[{{Kraft}(1967)}]{Kraft1967}
{Kraft} RP (1967) {Studies of Stellar Rotation. V. The Dependence of Rotation
  on Age among Solar-Type Stars}. \apj 150:551. \doi{10.1086/149359}

\bibitem[{{Krishnamurthi} et~al(1997){Krishnamurthi}, {Pinsonneault}, {Barnes},
  and {Sofia}}]{Krishnamurthi1997}
{Krishnamurthi} A, {Pinsonneault} MH, {Barnes} S, et~al (1997) {Theoretical
  Models of the Angular Momentum Evolution of Solar-Type Stars}. \apj
  480(1):303--323. \doi{10.1086/303958}

\bibitem[{{Landi Degl'Innocenti} and {Landolfi}(2004)}]{2004ASSL..307.....L}
{Landi Degl'Innocenti} E, {Landolfi} M (2004) {Polarization in Spectral Lines},
  vol 307. {Springer, Dordrecht}, \doi{10.1007/978-1-4020-2415-3}

\bibitem[{{Lanza} et~al(2008){Lanza}, {De Martino}, and
  {Rodon{\`o}}}]{Lanza+08}
{Lanza} AF, {De Martino} C, {Rodon{\`o}} M (2008) {Astrometric effects of
  solar-like magnetic activity in late-type stars and their relevance for the
  detection of extrasolar planets}. New Astronomy 13(2):77--84.
  \doi{10.1016/j.newast.2007.06.009},
  {\href{https://arxiv.org/abs/0706.2942}{{https://arxiv.org/abs/arXiv:0706.2942}}}
  {[astro-ph]}

\bibitem[{{Lehmann} et~al(2017){Lehmann}, {Jardine}, {Vidotto}, {Mackay},
  {See}, {Donati}, {Folsom}, {Jeffers}, {Marsden}, {Morin}, and
  {Petit}}]{Lehmann+17}
{Lehmann} LT, {Jardine} MM, {Vidotto} AA, et~al (2017) {The energy budget of
  stellar magnetic fields: comparing non-potential simulations and
  observations}. \mnras 466(1):L24--L28. \doi{10.1093/mnrasl/slw225},
  {\href{https://arxiv.org/abs/1610.08314}{{https://arxiv.org/abs/arXiv:1610.08314}}}
  {[astro-ph.SR]}

\bibitem[{{Lehmann} et~al(2019){Lehmann}, {Hussain}, {Jardine}, {Mackay}, and
  {Vidotto}}]{Lehmann+19}
{Lehmann} LT, {Hussain} GAJ, {Jardine} MM, et~al (2019) {Observing the
  simulations: applying ZDI to 3D non-potential magnetic field simulations}.
  \mnras 483(4):5246--5266. \doi{10.1093/mnras/sty3362},
  {\href{https://arxiv.org/abs/1811.03703}{{https://arxiv.org/abs/arXiv:1811.03703}}}
  {[astro-ph.SR]}

\bibitem[{{Li} et~al(2012){Li}, {Bi}, {Liu}, {Tian}, and {Shuai}}]{Li2012}
{Li} TD, {Bi} SL, {Liu} K, et~al (2012) {Stellar parameters and seismological
  analysis of the star 18 Scorpii}. \aap 546:A83.
  \doi{10.1051/0004-6361/201219063}

\bibitem[{{Lorenzo-Oliveira} et~al(2018){Lorenzo-Oliveira}, {Freitas},
  {Mel{\'e}ndez}, {Bedell}, {Ram{\'\i}rez}, {Bean}, {Asplund}, {Spina},
  {Dreizler}, {Alves-Brito}, and {Casagrande}}]{Lorenzo2018}
{Lorenzo-Oliveira} D, {Freitas} FC, {Mel{\'e}ndez} J, et~al (2018) {The Solar
  Twin Planet Search. The age-chromospheric activity relation}. \aap 619:A73.
  \doi{10.1051/0004-6361/201629294},
  {\href{https://arxiv.org/abs/1806.08014}{{https://arxiv.org/abs/arXiv:1806.08014}}}
  {[astro-ph.SR]}

\bibitem[{{Lu} et~al(2022){Lu}, {Curtis}, {Angus}, {David}, and
  {Hattori}}]{Lu2022}
{Lu} YL, {Curtis} JL, {Angus} R, et~al (2022) {Bridging the Gap-The
  Disappearance of the Intermediate Period Gap for Fully Convective Stars,
  Uncovered by New ZTF Rotation Periods}. \aj 164(6):251.
  \doi{10.3847/1538-3881/ac9bee},
  {\href{https://arxiv.org/abs/2210.06604}{{https://arxiv.org/abs/arXiv:2210.06604}}}
  {[astro-ph.SR]}

\bibitem[{{MacGregor} and {Brenner}(1991)}]{MacGregor1991}
{MacGregor} KB, {Brenner} M (1991) {Rotational Evolution of Solar-Type Stars.
  I. Main-Sequence Evolution}. \apj 376:204. \doi{10.1086/170269}

\bibitem[{{Mahadevan} et~al(2012){Mahadevan}, {Ramsey}, {Bender}, {Terrien},
  {Wright}, {Halverson}, {Hearty}, {Nelson}, {Burton}, {Redman}, {Osterman},
  {Diddams}, {Kasting}, {Endl}, and {Deshpande}}]{2012SPIE.8446E..1SM}
{Mahadevan} S, {Ramsey} L, {Bender} C, et~al (2012) {The habitable-zone planet
  finder: a stabilized fiber-fed NIR spectrograph for the Hobby-Eberly
  Telescope}. In: {McLean} IS, {Ramsay} SK, {Takami} H (eds) Ground-based and
  Airborne Instrumentation for Astronomy IV, p 84461S, \doi{10.1117/12.926102},
  \eprint{1209.1686}

\bibitem[{{Mamajek} and {Hillenbrand}(2008)}]{Mamajek2008}
{Mamajek} EE, {Hillenbrand} LA (2008) {Improved Age Estimation for Solar-Type
  Dwarfs Using Activity-Rotation Diagnostics}. \apj 687(2):1264--1293.
  \doi{10.1086/591785},
  {\href{https://arxiv.org/abs/0807.1686}{{https://arxiv.org/abs/arXiv:0807.1686}}}
  {[astro-ph]}

\bibitem[{{Marsden} et~al(2014){Marsden}, {Petit}, {Jeffers}, {Morin}, {Fares},
  {Reiners}, {do Nascimento}, {Auri{\`e}re}, {Bouvier}, {Carter}, {Catala},
  {Dintrans}, {Donati}, {Gastine}, {Jardine}, {Konstantinova-Antova}, {Lanoux},
  {Ligni{\`e}res}, {Morgenthaler}, {Ram{\`\i}rez-V{\`e}lez}, {Th{\'e}ado}, {Van
  Grootel}, and {BCool Collaboration}}]{2014MNRAS.444.3517M}
{Marsden} SC, {Petit} P, {Jeffers} SV, et~al (2014) {A BCool magnetic snapshot
  survey of solar-type stars}. \mnras 444(4):3517--3536.
  \doi{10.1093/mnras/stu1663},
  {\href{https://arxiv.org/abs/1311.3374}{{https://arxiv.org/abs/arXiv:1311.3374}}}
  {[astro-ph.SR]}

\bibitem[{{Masuda} et~al(2022){Masuda}, {Petigura}, and {Hall}}]{Masuda2022}
{Masuda} K, {Petigura} EA, {Hall} OJ (2022) {Inferring the rotation period
  distribution of stars from their projected rotation velocities and radii:
  Application to late-F/early-G Kepler stars}. \mnras 510(4):5623--5638.
  \doi{10.1093/mnras/stab3650},
  {\href{https://arxiv.org/abs/2112.07162}{{https://arxiv.org/abs/arXiv:2112.07162}}}
  {[astro-ph.SR]}

\bibitem[{{Matt} and {Pudritz}(2005)}]{Matt2005}
{Matt} S, {Pudritz} RE (2005) {Accretion-powered Stellar Winds as a Solution to
  the Stellar Angular Momentum Problem}. \apjl 632(2):L135--L138.
  \doi{10.1086/498066},
  {\href{https://arxiv.org/abs/astro-ph/0510060}{{https://arxiv.org/abs/arXiv:astro-ph/0510060}}}
  {[astro-ph]}

\bibitem[{{Matt} and {Pudritz}(2008)}]{Matt2008}
{Matt} S, {Pudritz} RE (2008) {Accretion-powered Stellar Winds. II. Numerical
  Solutions for Stellar Wind Torques}. \apj 678(2):1109--1118.
  \doi{10.1086/533428},
  {\href{https://arxiv.org/abs/0801.0436}{{https://arxiv.org/abs/arXiv:0801.0436}}}
  {[astro-ph]}

\bibitem[{{Matt} et~al(2015){Matt}, {Brun}, {Baraffe}, {Bouvier}, and
  {Chabrier}}]{Matt2015}
{Matt} SP, {Brun} AS, {Baraffe} I, et~al (2015) {The Mass-dependence of Angular
  Momentum Evolution in Sun-like Stars}. \apjl 799(2):L23.
  \doi{10.1088/2041-8205/799/2/L23},
  {\href{https://arxiv.org/abs/1412.4786}{{https://arxiv.org/abs/arXiv:1412.4786}}}
  {[astro-ph.SR]}

\bibitem[{{McQuillan} et~al(2014){McQuillan}, {Mazeh}, and
  {Aigrain}}]{McQuillan2014}
{McQuillan} A, {Mazeh} T, {Aigrain} S (2014) {Rotation Periods of 34,030 Kepler
  Main-sequence Stars: The Full Autocorrelation Sample}. \apjs 211(2):24.
  \doi{10.1088/0067-0049/211/2/24},
  {\href{https://arxiv.org/abs/1402.5694}{{https://arxiv.org/abs/arXiv:1402.5694}}}
  {[astro-ph.SR]}

\bibitem[{{Mestel}(1968)}]{Mestel1968}
{Mestel} L (1968) {Magnetic braking by a stellar wind-I}. \mnras 138:359.
  \doi{10.1093/mnras/138.3.359}

\bibitem[{{Metcalfe} et~al(2014){Metcalfe}, {Creevey}, {Do{\u{g}}an}, {Mathur},
  {Xu}, {Bedding}, {Chaplin}, {Christensen-Dalsgaard}, {Karoff}, {Trampedach},
  {Benomar}, {Brown}, {Buzasi}, {Campante}, {{\c{C}}elik}, {Cunha}, {Davies},
  {Deheuvels}, {Derekas}, {Di Mauro}, {Garc{\'\i}a}, {Guzik}, {Howe},
  {MacGregor}, {Mazumdar}, {Montalb{\'a}n}, {Monteiro}, {Salabert},
  {Serenelli}, {Stello}, {Ste\&{\c{s}}acute}, {licki}, {Suran}, {Y{\i}ld{\i}z},
  {Aksoy}, {Elsworth}, {Gruberbauer}, {Guenther}, {Lebreton}, {Molaverdikhani},
  {Pricopi}, {Simoniello}, and {White}}]{Metcalfe2014}
{Metcalfe} TS, {Creevey} OL, {Do{\u{g}}an} G, et~al (2014) {Properties of 42
  Solar-type Kepler Targets from the Asteroseismic Modeling Portal}. \apjs
  214(2):27. \doi{10.1088/0067-0049/214/2/27},
  {\href{https://arxiv.org/abs/1402.3614}{{https://arxiv.org/abs/arXiv:1402.3614}}}
  {[astro-ph.SR]}

\bibitem[{{Metcalfe} et~al(2016){Metcalfe}, {Egeland}, and {van
  Saders}}]{Metcalfe2016}
{Metcalfe} TS, {Egeland} R, {van Saders} J (2016) {Stellar Evidence That the
  Solar Dynamo May Be in Transition}. \apjl 826(1):L2.
  \doi{10.3847/2041-8205/826/1/L2},
  {\href{https://arxiv.org/abs/1606.01926}{{https://arxiv.org/abs/arXiv:1606.01926}}}
  {[astro-ph.SR]}

\bibitem[{{Metcalfe} et~al(2023){Metcalfe}, {Strassmeier}, {Ilyin}, {van
  Saders}, {Ayres}, {Finley}, {Kochukhov}, {Petit}, {See}, {Stassun},
  {Jeffers}, {Marsden}, {Morin}, and {Vidotto}}]{Metcalfe2023}
{Metcalfe} TS, {Strassmeier} KG, {Ilyin} IV, et~al (2023) {Constraints on
  Magnetic Braking from the G8 Dwarf Stars 61 UMa and {\ensuremath{\tau}} Cet}.
  \apjl 948(1):L6. \doi{10.3847/2041-8213/acce38},
  {\href{https://arxiv.org/abs/2304.09896}{{https://arxiv.org/abs/arXiv:2304.09896}}}
  {[astro-ph.SR]}

\bibitem[{{Meunier} and {Lagrange}(2019)}]{Meunier+19b}
{Meunier} N, {Lagrange} AM (2019) {Activity time series of old stars from late
  F to early K. II. Radial velocity jitter and exoplanet detectability}. \aap
  628:A125. \doi{10.1051/0004-6361/201935347},
  {\href{https://arxiv.org/abs/1909.02969}{{https://arxiv.org/abs/arXiv:1909.02969}}}
  {[astro-ph.SR]}

\bibitem[{{Meunier} et~al(2019){Meunier}, {Lagrange}, {Boulet}, and
  {Borgniet}}]{Meunier+19a}
{Meunier} N, {Lagrange} AM, {Boulet} T, et~al (2019) {Activity time series of
  old stars from late F to early K. I. Simulating radial velocity, astrometry,
  photometry, and chromospheric emission}. \aap 627:A56.
  \doi{10.1051/0004-6361/201834796},
  {\href{https://arxiv.org/abs/1904.01437}{{https://arxiv.org/abs/arXiv:1904.01437}}}
  {[astro-ph.SR]}

\bibitem[{{Meunier} et~al(2020){Meunier}, {Lagrange}, and
  {Borgniet}}]{Meunier+20}
{Meunier} N, {Lagrange} AM, {Borgniet} S (2020) {Activity time series of old
  stars from late F to early K. V. Effect on exoplanet detectability with
  high-precision astrometry}. \aap 644:A77. \doi{10.1051/0004-6361/202038710},
  {\href{https://arxiv.org/abs/2011.02158}{{https://arxiv.org/abs/arXiv:2011.02158}}}
  {[astro-ph.SR]}

\bibitem[{{N{\`e}mec} et~al(2022){N{\`e}mec}, {Shapiro}, {I{\c{s}}{\i}k},
  {Sowmya}, {Solanki}, {Krivova}, {Cameron}, and {Gizon}}]{Nemec+22}
{N{\`e}mec} NE, {Shapiro} AI, {I{\c{s}}{\i}k} E, et~al (2022) {Faculae Cancel
  out on the Surfaces of Active Suns}. \apjl 934(2):L23.
  \doi{10.3847/2041-8213/ac8155},
  {\href{https://arxiv.org/abs/2207.06816}{{https://arxiv.org/abs/arXiv:2207.06816}}}
  {[astro-ph.SR]}

\bibitem[{{N{\`e}mec} et~al(2023){N{\`e}mec}, {Shapiro}, {I{\c{s}}{\i}k},
  {Solanki}, and {Reinhold}}]{Nemec+23}
{N{\`e}mec} NE, {Shapiro} AI, {I{\c{s}}{\i}k} E, et~al (2023) {Forward
  modelling of brightness variations in Sun-like stars. II. Light curves and
  variability}. \aap 672:A138. \doi{10.1051/0004-6361/202244412},
  {\href{https://arxiv.org/abs/2303.03040}{{https://arxiv.org/abs/arXiv:2303.03040}}}
  {[astro-ph.SR]}

\bibitem[{{Noyes} et~al(1984){Noyes}, {Hartmann}, {Baliunas}, {Duncan}, and
  {Vaughan}}]{Noyes1984}
{Noyes} RW, {Hartmann} LW, {Baliunas} SL, et~al (1984) {Rotation, convection,
  and magnetic activity in lower main-sequence stars.} \apj 279:763--777.
  \doi{10.1086/161945}

\bibitem[{{Oglethorpe} and {Garaud}(2013)}]{Oglethorpe2013}
{Oglethorpe} RLF, {Garaud} P (2013) {Spin-down Dynamics of Magnetized
  Solar-type Stars}. \apj 778(2):166. \doi{10.1088/0004-637X/778/2/166},
  {\href{https://arxiv.org/abs/1401.0932}{{https://arxiv.org/abs/arXiv:1401.0932}}}
  {[astro-ph.SR]}

\bibitem[{{Pevtsov} et~al(2003){Pevtsov}, {Fisher}, {Acton}, {Longcope},
  {Johns-Krull}, {Kankelborg}, and {Metcalf}}]{2003ApJ...598.1387P}
{Pevtsov} AA, {Fisher} GH, {Acton} LW, et~al (2003) {The Relationship Between
  X-Ray Radiance and Magnetic Flux}. \apj 598:1387--1391. \doi{10.1086/378944}

\bibitem[{{Piskunov} et~al(2011){Piskunov}, {Snik}, {Dolgopolov}, {Kochukhov},
  {Rodenhuis}, {Valenti}, {Jeffers}, {Makaganiuk}, {Johns-Krull}, {Stempels},
  and {Keller}}]{Piskunov11}
{Piskunov} N, {Snik} F, {Dolgopolov} A, et~al (2011) {HARPSpol {\textemdash}
  The New Polarimetric Mode for HARPS}. The Messenger 143:7--10

\bibitem[{{Pizzolato} et~al(2003){Pizzolato}, {Maggio}, {Micela}, {Sciortino},
  and {Ventura}}]{Pizzolato2003}
{Pizzolato} N, {Maggio} A, {Micela} G, et~al (2003) {The stellar
  activity-rotation relationship revisited: Dependence of saturated and
  non-saturated X-ray emission regimes on stellar mass for late-type dwarfs}.
  \aap 397:147--157. \doi{10.1051/0004-6361:20021560}

\bibitem[{{Quirrenbach} et~al(2014){Quirrenbach}, {Amado}, {Caballero},
  {Mundt}, {Reiners}, {Ribas}, {Seifert}, {Abril}, {Aceituno},
  {Alonso-Floriano}, {Ammler-von Eiff}, {Antona Jim{\'e}nez},
  {Anwand-Heerwart}, {Azzaro}, {Bauer}, {Barrado}, {Becerril}, {B{\'e}jar},
  {Ben{\'\i}tez}, {Berdi{\~n}as}, {C{\'a}rdenas}, {Casal}, {Claret},
  {Colom{\'e}}, {Cort{\'e}s-Contreras}, {Czesla}, {Doellinger}, {Dreizler},
  {Feiz}, {Fern{\'a}ndez}, {Galad{\'\i}}, {G{\'a}lvez-Ortiz},
  {Garc{\'\i}a-Piquer}, {Garc{\'\i}a-Vargas}, {Garrido}, {Gesa}, {G{\'o}mez
  Galera}, {Gonz{\'a}lez {\'A}lvarez}, {Gonz{\'a}lez Hern{\'a}ndez},
  {Gr{\"o}zinger}, {Gu{\`a}rdia}, {Guenther}, {de Guindos},
  {Guti{\'e}rrez-Soto}, {Hagen}, {Hatzes}, {Hauschildt}, {Helmling}, {Henning},
  {Hermann}, {Hern{\'a}ndez Casta{\~n}o}, {Herrero}, {Hidalgo}, {Holgado},
  {Huber}, {Huber}, {Jeffers}, {Joergens}, {de Juan}, {Kehr}, {Klein},
  {K{\"u}rster}, {Lamert}, {Lalitha}, {Laun}, {Lemke}, {Lenzen}, {L{\'o}pez del
  Fresno}, {L{\'o}pez Mart{\'\i}}, {L{\'o}pez-Santiago}, {Mall}, {Mandel},
  {Mart{\'\i}n}, {Mart{\'\i}n-Ruiz}, {Mart{\'\i}nez-Rodr{\'\i}guez}, {Marvin},
  {Mathar}, {Mirabet}, {Montes}, {Morales Mu{\~n}oz}, {Moya}, {Naranjo},
  {Ofir}, {Oreiro}, {Pall{\'e}}, {Panduro}, {Passegger}, {P{\'e}rez-Calpena},
  {P{\'e}rez Medialdea}, {Perger}, {Pluto}, {Ram{\'o}n}, {Rebolo}, {Redondo},
  {Reffert}, {Reinhardt}, {Rhode}, {Rix}, {Rodler}, {Rodr{\'\i}guez},
  {Rodr{\'\i}guez-L{\'o}pez}, {Rodr{\'\i}guez-P{\'e}rez}, {Rohloff}, {Rosich},
  {S{\'a}nchez-Blanco}, {S{\'a}nchez Carrasco}, {Sanz-Forcada}, {Sarmiento},
  {Sch{\"a}fer}, {Schiller}, {Schmidt}, {Schmitt}, {Solano}, {Stahl}, {Storz},
  {St{\"u}rmer}, {Su{\'a}rez}, {Ulbrich}, {Veredas}, {Wagner}, {Winkler},
  {Zapatero Osorio}, {Zechmeister}, {Abell{\'a}n de Paco},
  {Anglada-Escud{\'e}}, {del Burgo}, {Klutsch}, {Lizon}, {L{\'o}pez-Morales},
  {Morales}, {Perryman}, {Tulloch}, and {Xu}}]{carmenes}
{Quirrenbach} A, {Amado} PJ, {Caballero} JA, et~al (2014) {CARMENES instrument
  overview}. In: {Ramsay} SK, {McLean} IS, {Takami} H (eds) Ground-based and
  Airborne Instrumentation for Astronomy V, p 91471F, \doi{10.1117/12.2056453}

\bibitem[{{Radick} et~al(1998){Radick}, {Lockwood}, {Skiff}, and
  {Baliunas}}]{Radick+18}
{Radick} RR, {Lockwood} GW, {Skiff} BA, et~al (1998) {Patterns of Variation
  among Sun-like Stars}. \apjs 118(1):239--258. \doi{10.1086/313135}

\bibitem[{{Reiners}(2012)}]{2012LRSP....9....1R}
{Reiners} A (2012) {Observations of Cool-Star Magnetic Fields}. Living Reviews
  in Solar Physics 9:1.
  {\href{https://arxiv.org/abs/1203.0241}{{https://arxiv.org/abs/arXiv:1203.0241}}}
  {[astro-ph.SR]}

\bibitem[{{Reiners} et~al(2014){Reiners}, {Sch{\"u}ssler}, and
  {Passegger}}]{2014ApJ...794..144R}
{Reiners} A, {Sch{\"u}ssler} M, {Passegger} VM (2014) {Generalized
  Investigation of the Rotation-Activity Relation: Favoring Rotation Period
  instead of Rossby Number}. \apj 794(2):144.
  \doi{10.1088/0004-637X/794/2/144},
  {\href{https://arxiv.org/abs/1408.6175}{{https://arxiv.org/abs/arXiv:1408.6175}}}
  {[astro-ph.SR]}

\bibitem[{{Reiners} et~al(2022){Reiners}, {Shulyak}, {K{\"a}pyl{\"a}}, {Ribas},
  {Nagel}, {Zechmeister}, {Caballero}, {Shan}, {Fuhrmeister}, {Quirrenbach},
  {Amado}, {Montes}, {Jeffers}, {Azzaro}, {B{\'e}jar}, {Chaturvedi}, {Henning},
  {K{\"u}rster}, and {Pall{\'e}}}]{2022A&A...662A..41R}
{Reiners} A, {Shulyak} D, {K{\"a}pyl{\"a}} PJ, et~al (2022) {Magnetism,
  rotation, and nonthermal emission in cool stars. Average magnetic field
  measurements in 292 M dwarfs}. \aap 662:A41.
  \doi{10.1051/0004-6361/202243251},
  {\href{https://arxiv.org/abs/2204.00342}{{https://arxiv.org/abs/arXiv:2204.00342}}}
  {[astro-ph.SR]}

\bibitem[{{Reinhold} et~al(2020){Reinhold}, {Shapiro}, {Solanki}, {Montet},
  {Krivova}, {Cameron}, and {Amazo-G{\'o}mez}}]{Reinhold+20}
{Reinhold} T, {Shapiro} AI, {Solanki} SK, et~al (2020) {The Sun is less active
  than other solar-like stars}. Science 368(6490):518--521.
  \doi{10.1126/science.aay3821},
  {\href{https://arxiv.org/abs/2005.01401}{{https://arxiv.org/abs/arXiv:2005.01401}}}
  {[astro-ph.SR]}

\bibitem[{{R{\'e}ville} et~al(2015){R{\'e}ville}, {Brun}, {Matt}, {Strugarek},
  and {Pinto}}]{Reville2015}
{R{\'e}ville} V, {Brun} AS, {Matt} SP, et~al (2015) {The Effect of Magnetic
  Topology on Thermally Driven Wind: Toward a General Formulation of the
  Braking Law}. \apj 798(2):116. \doi{10.1088/0004-637X/798/2/116},
  {\href{https://arxiv.org/abs/1410.8746}{{https://arxiv.org/abs/arXiv:1410.8746}}}
  {[astro-ph.SR]}

\bibitem[{{Roquette} et~al(2021){Roquette}, {Matt}, {Winter}, {Amard}, and
  {Stasevic}}]{Roquette2021}
{Roquette} J, {Matt} SP, {Winter} AJ, et~al (2021) {The influence of the
  environment on the spin evolution of low-mass stars - I. External
  photoevaporation of circumstellar discs}. \mnras 508(3):3710--3729.
  \doi{10.1093/mnras/stab2772},
  {\href{https://arxiv.org/abs/2109.10296}{{https://arxiv.org/abs/arXiv:2109.10296}}}
  {[astro-ph.SR]}

\bibitem[{{Saar}(1988)}]{1988ApJ...324..441S}
{Saar} SH (1988) {Improved Methods for the Measurement and Analysis of Stellar
  Magnetic Fields}. \apj 324:441. \doi{10.1086/165907}

\bibitem[{{Schmitt} and {Liefke}(2004)}]{SchmittLiefke2004}
{Schmitt} JHMM, {Liefke} C (2004) {NEXXUS: A comprehensive ROSAT survey of
  coronal X-ray emission among nearby solar-like stars}. \aap 417:651--665.
  \doi{10.1051/0004-6361:20030495},
  {\href{https://arxiv.org/abs/astro-ph/0308510}{{https://arxiv.org/abs/arXiv:astro-ph/0308510}}}
  {[astro-ph]}

\bibitem[{{Schrijver}(1987)}]{1987A&A...180..241S}
{Schrijver} CJ (1987) {Solar active regions - Radiative intensities and
  large-scale parameters of the magnetic field}. \aap 180(1-2):241--252

\bibitem[{{Schrijver} and {Title}(2001)}]{schrijver01}
{Schrijver} CJ, {Title} AM (2001) {On the Formation of Polar Spots in Sun-like
  Stars}. \apj 551(2):1099--1106. \doi{10.1086/320237}

\bibitem[{{Schrijver} and {Zwaan}(2000)}]{schrijver00}
{Schrijver} CJ, {Zwaan} C (2000) {Solar and Stellar Magnetic Activity}.
  {Cambridge University Press, Cambridge}

\bibitem[{{Sch\"ussler} and {Solanki}(1992)}]{schuessler92}
{Sch\"ussler} M, {Solanki} SK (1992) {Why rapid rotators have polar spots.}
  \aap 264:L13--L16

\bibitem[{{Sch\"ussler} et~al(1996){Sch\"ussler}, {Caligari}, {Ferriz-Mas},
  {Solanki}, and {Stix}}]{schuessler96}
{Sch\"ussler} M, {Caligari} P, {Ferriz-Mas} A, et~al (1996) {Distribution of
  starspots on cool stars. I. Young and main sequence stars of 1M$_{sun}$\_.}
  \aap 314:503--512

\bibitem[{{See} et~al(2015){See}, {Jardine}, {Vidotto}, {Donati}, {Folsom},
  {Boro Saikia}, {Bouvier}, {Fares}, {Gregory}, {Hussain}, {Jeffers},
  {Marsden}, {Morin}, {Moutou}, {do Nascimento}, {Petit}, {Ros{\'e}n}, and
  {Waite}}]{See+15}
{See} V, {Jardine} M, {Vidotto} AA, et~al (2015) {The energy budget of stellar
  magnetic fields}. \mnras 453(4):4301--4310. \doi{10.1093/mnras/stv1925},
  {\href{https://arxiv.org/abs/1508.01403}{{https://arxiv.org/abs/arXiv:1508.01403}}}
  {[astro-ph.SR]}

\bibitem[{{See} et~al(2019){See}, {Matt}, {Finley}, {Folsom}, {Boro Saikia},
  {Donati}, {Fares}, {H{\'e}brard}, {Jardine}, {Jeffers}, {Marsden}, {Mengel},
  {Morin}, {Petit}, {Vidotto}, {Waite}, and {BCool Collaboration}}]{See2019}
{See} V, {Matt} SP, {Finley} AJ, et~al (2019) {Do Non-dipolar Magnetic Fields
  Contribute to Spin-down Torques?} \apj 886(2):120.
  \doi{10.3847/1538-4357/ab46b2},
  {\href{https://arxiv.org/abs/1910.02129}{{https://arxiv.org/abs/arXiv:1910.02129}}}
  {[astro-ph.SR]}

\bibitem[{{{\c{S}}enavc{\i}} et~al(2021){{\c{S}}enavc{\i}},
  {K{\i}l{\i}{\c{c}}o{\u{g}}lu}, {I{\c{s}}{\i}k}, {Hussain}, {Montes}, {Bahar},
  and {Solanki}}]{Senavci+21}
{{\c{S}}enavc{\i}} HV, {K{\i}l{\i}{\c{c}}o{\u{g}}lu} T, {I{\c{s}}{\i}k} E,
  et~al (2021) {Observing and modelling the young solar analogue EK Draconis:
  starspot distribution, elemental abundances, and evolutionary status}. \mnras
  502(3):3343--3356. \doi{10.1093/mnras/stab199},
  {\href{https://arxiv.org/abs/2101.07248}{{https://arxiv.org/abs/arXiv:2101.07248}}}
  {[astro-ph.SR]}

\bibitem[{{Shapiro} et~al(2014){Shapiro}, {Solanki}, {Krivova}, {Schmutz},
  {Ball}, {Knaack}, {Rozanov}, and {Unruh}}]{Shapiro+14}
{Shapiro} AI, {Solanki} SK, {Krivova} NA, et~al (2014) {Variability of Sun-like
  stars: reproducing observed photometric trends}. \aap 569:A38.
  \doi{10.1051/0004-6361/201323086},
  {\href{https://arxiv.org/abs/1406.2383}{{https://arxiv.org/abs/arXiv:1406.2383}}}
  {[astro-ph.SR]}

\bibitem[{{Shapiro} et~al(2021){Shapiro}, {Solanki}, and
  {Krivova}}]{Shapiro+21}
{Shapiro} AI, {Solanki} SK, {Krivova} NA (2021) {Predictions of Astrometric
  Jitter for Sun-like Stars. I. The Model and Its Application to the Sun as
  Seen from the Ecliptic}. \apj 908(2):223. \doi{10.3847/1538-4357/abd630},
  {\href{https://arxiv.org/abs/2012.12312}{{https://arxiv.org/abs/arXiv:2012.12312}}}
  {[astro-ph.SR]}

\bibitem[{{Shu} et~al(1994){Shu}, {Najita}, {Ostriker}, {Wilkin}, {Ruden}, and
  {Lizano}}]{Shu1994}
{Shu} F, {Najita} J, {Ostriker} E, et~al (1994) {Magnetocentrifugally Driven
  Flows from Young Stars and Disks. I. A Generalized Model}. \apj 429:781.
  \doi{10.1086/174363}

\bibitem[{{Skumanich}(1972)}]{Skumanich1972}
{Skumanich} A (1972) {Time Scales for Ca II Emission Decay, Rotational Braking,
  and Lithium Depletion}. \apj 171:565. \doi{10.1086/151310}

\bibitem[{{Snik} et~al(2008){Snik}, {Jeffers}, {Keller}, {Piskunov},
  {Kochukhov}, {Valenti}, and {Johns-Krull}}]{Snik08}
{Snik} F, {Jeffers} S, {Keller} C, et~al (2008) {The upgrade of HARPS to a
  full-Stokes high-resolution spectropolarimeter}. In: {McLean} IS, {Casali} MM
  (eds) Ground-based and Airborne Instrumentation for Astronomy II, p 70140O,
  \doi{10.1117/12.787393}

\bibitem[{{Solanki}(2002)}]{solanki02}
{Solanki} SK (2002) {The magnetic structure of sunspots and starspots}.
  Astronomische Nachrichten 323:165--177.
  \doi{10.1002/1521-3994(200208)323:3/4<165::AID-ASNA165>3.0.CO;2-U}

\bibitem[{{Solanki}(2003)}]{2003A&ARv..11..153S}
{Solanki} SK (2003) {Sunspots: An overview}. \aapr 11(2-3):153--286.
  \doi{10.1007/s00159-003-0018-4}

\bibitem[{{Solanki} et~al(2006){Solanki}, {Inhester}, and
  {Sch{\"u}ssler}}]{2006RPPh...69..563S}
{Solanki} SK, {Inhester} B, {Sch{\"u}ssler} M (2006) {The solar magnetic
  field}. Reports on Progress in Physics 69(3):563--668.
  \doi{10.1088/0034-4885/69/3/R02},
  {\href{https://arxiv.org/abs/1008.0771}{{https://arxiv.org/abs/arXiv:1008.0771}}}
  {[astro-ph.SR]}

\bibitem[{{Somers} and {Pinsonneault}(2016)}]{Somers2016}
{Somers} G, {Pinsonneault} MH (2016) {Lithium Depletion is a Strong Test of
  Core-envelope Recoupling}. \apj 829(1):32. \doi{10.3847/0004-637X/829/1/32},
  {\href{https://arxiv.org/abs/1606.00004}{{https://arxiv.org/abs/arXiv:1606.00004}}}
  {[astro-ph.SR]}

\bibitem[{{Somers} et~al(2017){Somers}, {Stauffer}, {Rebull}, {Cody}, and
  {Pinsonneault}}]{Somers2017}
{Somers} G, {Stauffer} J, {Rebull} L, et~al (2017) {M Dwarf Rotation from the
  K2 Young Clusters to the Field. I. A Mass-Rotation Correlation at 10 Myr}.
  \apj 850(2):134. \doi{10.3847/1538-4357/aa93ed},
  {\href{https://arxiv.org/abs/1710.07638}{{https://arxiv.org/abs/arXiv:1710.07638}}}
  {[astro-ph.SR]}

\bibitem[{{Sowmya} et~al(2021{\natexlab{a}}){Sowmya}, {N{\`e}mec}, {Shapiro},
  {I{\c{s}}{\i}k}, {Witzke}, {Mints}, {Krivova}, and {Solanki}}]{Sowmya2021}
{Sowmya} K, {N{\`e}mec} NE, {Shapiro} AI, et~al (2021{\natexlab{a}})
  {Predictions of Astrometric Jitter for Sun-like Stars. II. Dependence on
  Inclination, Metallicity, and Active-region Nesting}. \apj 919(2):94.
  \doi{10.3847/1538-4357/ac111b},
  {\href{https://arxiv.org/abs/2107.01493}{{https://arxiv.org/abs/arXiv:2107.01493}}}
  {[astro-ph.SR]}

\bibitem[{{Sowmya} et~al(2021{\natexlab{b}}){Sowmya}, {Shapiro}, {Witzke},
  {N{\`e}mec}, {Chatzistergos}, {Yeo}, {Krivova}, and {Solanki}}]{Sowmya+21}
{Sowmya} K, {Shapiro} AI, {Witzke} V, et~al (2021{\natexlab{b}}) {Modeling
  Stellar Ca II H and K Emission Variations. I. Effect of Inclination on the
  S-index}. \apj 914(1):21. \doi{10.3847/1538-4357/abf247},
  {\href{https://arxiv.org/abs/2103.13893}{{https://arxiv.org/abs/arXiv:2103.13893}}}
  {[astro-ph.SR]}

\bibitem[{{Sowmya} et~al(2022){Sowmya}, {N{\`e}mec}, {Shapiro},
  {I{\c{s}}{\i}k}, {Krivova}, and {Solanki}}]{Sowmya+22}
{Sowmya} K, {N{\`e}mec} NE, {Shapiro} AI, et~al (2022) {Predictions of
  Astrometric Jitter for Sun-like Stars. III. Fast Rotators}. \apj 934(2):146.
  \doi{10.3847/1538-4357/ac79b3},
  {\href{https://arxiv.org/abs/2206.07702}{{https://arxiv.org/abs/arXiv:2206.07702}}}
  {[astro-ph.SR]}

\bibitem[{{Spada} and {Lanzafame}(2020)}]{Spada2020}
{Spada} F, {Lanzafame} AC (2020) {Competing effect of wind braking and interior
  coupling in the rotational evolution of solar-like stars}. \aap 636:A76.
  \doi{10.1051/0004-6361/201936384},
  {\href{https://arxiv.org/abs/1908.00345}{{https://arxiv.org/abs/arXiv:1908.00345}}}
  {[astro-ph.SR]}

\bibitem[{{Strassmeier}(2009)}]{Strassmeier09}
{Strassmeier} KG (2009) {Starspots}. \aapr 17(3):251--308.
  \doi{10.1007/s00159-009-0020-6}

\bibitem[{{Strassmeier} and {Rice}(1998)}]{strassrice98}
{Strassmeier} KG, {Rice} JB (1998) {Doppler imaging of stellar surface
  structure. VI. HD 129333 = EK Draconis: a stellar analog of the active young
  Sun}. \aap 330:685--695

\bibitem[{{Strassmeier} et~al(2015){Strassmeier}, {Ilyin}, {J{\"a}rvinen},
  {Weber}, {Woche}, {Barnes}, {Bauer}, {Beckert}, {Bittner}, {Bredthauer},
  {Carroll}, {Denker}, {Dionies}, {DiVarano}, {D{\"o}scher}, {Fechner},
  {Feuerstein}, {Granzer}, {Hahn}, {Harnisch}, {Hofmann}, {Lesser}, {Paschke},
  {Pankratow}, {Plank}, {Pl{\"u}schke}, {Popow}, and
  {Sablowski}}]{2015AN....336..324S}
{Strassmeier} KG, {Ilyin} I, {J{\"a}rvinen} A, et~al (2015) {PEPSI: The
  high-resolution {\'e}chelle spectrograph and polarimeter for the Large
  Binocular Telescope}. Astronomische Nachrichten 336(4):324.
  \doi{10.1002/asna.201512172},
  {\href{https://arxiv.org/abs/1505.06492}{{https://arxiv.org/abs/arXiv:1505.06492}}}
  {[astro-ph.IM]}

\bibitem[{{van Saders} and {Pinsonneault}(2013)}]{vanSaders2013}
{van Saders} JL, {Pinsonneault} MH (2013) {Fast Star, Slow Star; Old Star,
  Young Star: Subgiant Rotation as a Population and Stellar Physics
  Diagnostic}. \apj 776(2):67. \doi{10.1088/0004-637X/776/2/67},
  {\href{https://arxiv.org/abs/1306.3701}{{https://arxiv.org/abs/arXiv:1306.3701}}}
  {[astro-ph.SR]}

\bibitem[{{van Saders} et~al(2016){van Saders}, {Ceillier}, {Metcalfe}, {Silva
  Aguirre}, {Pinsonneault}, {Garc{\'\i}a}, {Mathur}, and
  {Davies}}]{vanSaders2016}
{van Saders} JL, {Ceillier} T, {Metcalfe} TS, et~al (2016) {Weakened magnetic
  braking as the origin of anomalously rapid rotation in old field stars}. \nat
  529(7585):181--184. \doi{10.1038/nature16168},
  {\href{https://arxiv.org/abs/1601.02631}{{https://arxiv.org/abs/arXiv:1601.02631}}}
  {[astro-ph.SR]}

\bibitem[{{van Saders} et~al(2019){van Saders}, {Pinsonneault}, and
  {Barbieri}}]{vanSaders2019}
{van Saders} JL, {Pinsonneault} MH, {Barbieri} M (2019) {Forward Modeling of
  the Kepler Stellar Rotation Period Distribution: Interpreting Periods from
  Mixed and Biased Stellar Populations}. \apj 872(2):128.
  \doi{10.3847/1538-4357/aafafe},
  {\href{https://arxiv.org/abs/1803.04971}{{https://arxiv.org/abs/arXiv:1803.04971}}}
  {[astro-ph.SR]}

\bibitem[{{Vaughan} et~al(1978){Vaughan}, {Preston}, and
  {Wilson}}]{Vaughan1978}
{Vaughan} AH, {Preston} GW, {Wilson} OC (1978) {Flux measurements of Ca II and
  K emission.} \pasp 90:267--274. \doi{10.1086/130324}

\bibitem[{{Vidotto}(2016)}]{Vidotto16}
{Vidotto} AA (2016) {The magnetic field vector of the Sun-as-a-star}. \mnras
  459(2):1533--1542. \doi{10.1093/mnras/stw758},
  {\href{https://arxiv.org/abs/1603.09226}{{https://arxiv.org/abs/arXiv:1603.09226}}}
  {[astro-ph.SR]}

\bibitem[{{Vidotto} et~al(2014){Vidotto}, {Gregory}, {Jardine}, {Donati},
  {Petit}, {Morin}, {Folsom}, {Bouvier}, {Cameron}, {Hussain}, {Marsden},
  {Waite}, {Fares}, {Jeffers}, and {do Nascimento}}]{2014MNRAS.441.2361V}
{Vidotto} AA, {Gregory} SG, {Jardine} M, et~al (2014) {Stellar magnetism:
  empirical trends with age and rotation}. \mnras 441(3):2361--2374.
  \doi{10.1093/mnras/stu728},
  {\href{https://arxiv.org/abs/1404.2733}{{https://arxiv.org/abs/arXiv:1404.2733}}}
  {[astro-ph.SR]}

\bibitem[{{Vilhu}(1984)}]{Vilhu1984}
{Vilhu} O (1984) {The nature of magnetic activity in lower main sequence
  stars.} \aap 133:117--126

\bibitem[{{Weber} and {Davis}(1967)}]{Weber1967}
{Weber} EJ, {Davis} JLeverett (1967) {The Angular Momentum of the Solar Wind}.
  \apj 148:217--227. \doi{10.1086/149138}

\bibitem[{{Weber} et~al(2023){Weber}, {Schunker}, {Jouve}, and
  {I{\c{s}}{\i}k}}]{Weber+23}
{Weber} MA, {Schunker} H, {Jouve} L, et~al (2023) {Understanding Active Region
  Emergence and Origins on the Sun and Other Cool Stars}. arXiv e-prints
  arXiv:2306.06536. \doi{10.48550/arXiv.2306.06536},
  {\href{https://arxiv.org/abs/2306.06536}{{https://arxiv.org/abs/arXiv:2306.06536}}}
  {[astro-ph.SR]}

\bibitem[{{Wilson}(1968)}]{Wilson1968}
{Wilson} OC (1968) {Flux Measurements at the Centers of Stellar H- and
  K-Lines}. \apj 153:221. \doi{10.1086/149652}

\bibitem[{{Wood} et~al(2005){Wood}, {M{\"u}ller}, {Zank}, {Linsky}, and
  {Redfield}}]{Wood2005}
{Wood} BE, {M{\"u}ller} HR, {Zank} GP, et~al (2005) {New Mass-Loss Measurements
  from Astrospheric Ly{\ensuremath{\alpha}} Absorption}. \apjl
  628(2):L143--L146. \doi{10.1086/432716},
  {\href{https://arxiv.org/abs/astro-ph/0506401}{{https://arxiv.org/abs/arXiv:astro-ph/0506401}}}
  {[astro-ph]}

\bibitem[{{Wood} et~al(2021){Wood}, {M{\"u}ller}, {Redfield}, {Konow},
  {Vannier}, {Linsky}, {Youngblood}, {Vidotto}, {Jardine},
  {Alvarado-G{\'o}mez}, and {Drake}}]{Wood2021}
{Wood} BE, {M{\"u}ller} HR, {Redfield} S, et~al (2021) {New Observational
  Constraints on the Winds of M dwarf Stars}. \apj 915(1):37.
  \doi{10.3847/1538-4357/abfda5},
  {\href{https://arxiv.org/abs/2105.00019}{{https://arxiv.org/abs/arXiv:2105.00019}}}
  {[astro-ph.SR]}

\bibitem[{{Wright} et~al(2011){Wright}, {Drake}, {Mamajek}, and
  {Henry}}]{Wright2011}
{Wright} NJ, {Drake} JJ, {Mamajek} EE, et~al (2011) {The
  Stellar-activity-Rotation Relationship and the Evolution of Stellar Dynamos}.
  \apj 743(1):48. \doi{10.1088/0004-637X/743/1/48},
  {\href{https://arxiv.org/abs/1109.4634}{{https://arxiv.org/abs/arXiv:1109.4634}}}
  {[astro-ph.SR]}

\bibitem[{{Yadav} et~al(2015){Yadav}, {Christensen}, {Morin}, {Gastine},
  {Reiners}, {Poppenhaeger}, and {Wolk}}]{yadav15}
{Yadav} RK, {Christensen} UR, {Morin} J, et~al (2015) {Explaining the
  Coexistence of Large-scale and Small-scale Magnetic Fields in Fully
  Convective Stars}. \apjl 813(2):L31. \doi{10.1088/2041-8205/813/2/L31},
  {\href{https://arxiv.org/abs/1510.05541}{{https://arxiv.org/abs/arXiv:1510.05541}}}
  {[astro-ph.SR]}

\end{thebibliography}

\end{document}